\documentclass[12pt]{article}
\pdfoutput=1 

\usepackage{jheppub} 

\usepackage{graphicx}
\usepackage{caption}
\usepackage{epsfig}
\usepackage{epstopdf}
\usepackage{amsmath}
\usepackage{amssymb}
\usepackage{mathrsfs}
\usepackage{scalerel,stackengine}
\usepackage{mathdesign}
\usepackage{amsfonts}
\usepackage{bbold}
\usepackage{bbm}
\usepackage{dsfont}
\usepackage{dcolumn}
\usepackage{bm}
\usepackage{float}
\usepackage{comment}
\usepackage{hyperref}
\usepackage{xcolor}

\usepackage{wasysym}
\usepackage{mathtools}
\makeatletter \@addtoreset{equation}{section}

\newcommand\Shahin[1]{\textcolor{red}{ [ShJ:\,#1]}}
\newcommand\vahid[1]{\textcolor{magenta}{ [Vahid:\,#1]}}
\newcommand\ali[1]{\textcolor{cyan}{ [Ali:\,#1]}}
\newcommand{\cA}{\mathcal{A}}
\newcommand{\cF}{\mathcal{F}}
\newcommand{\cL}{\mathcal{L}}
\newcommand{\cQ}{\mathcal{Q}}
\newcommand{\bcQ}{\boldsymbol{\mathcal{Q}}}
\newcommand{\bz}{\bar{z}}

\newcommand{\bQ}{\boldsymbol{Q}}
\newcommand{\bOmega}{\boldsymbol{\Omega}}
\newcommand{\balr}{\boldsymbol{\alpha}^R}\newcommand{\ball}{\boldsymbol{\alpha}^L}\newcommand{\bpir}{\boldsymbol{\pi}^R}\newcommand{\bpil}{\boldsymbol{\pi}^L}
\newcommand{\bLr}{\boldsymbol{\mathcal{L}}^R}\newcommand{\bLl}{\boldsymbol{\mathcal{L}}^L}\newcommand{\bcL}{\boldsymbol{\mathcal{L}}}
\newcommand{\df}{\mathbbmss{d}}
\newcommand{\di}{\mathbbmss{i}}

\newcommand{\ee}{\end{equation}}\newcommand{\be}{\begin{equation}}

\graphicspath{{Graphics/}}

\begin{document}

\preprint{IPM/P-2018/030}

\title{\textbf{ Soft Charges and Electric-Magnetic Duality}}

\author[]{V. Hosseinzadeh, A. Seraj, M.M. Sheikh-Jabbari}

\affiliation[]{School of Physics, Institute for Research in Fundamental Sciences (IPM),\\P.O.Box 19395-5531,Tehran, IRAN}
\emailAdd{v.hosseinzadeh@ipm.ir, ali\_seraj@ipm.ir, jabbari@theory.ipm.ac.ir}

\abstract{The main focus of this work is to study magnetic soft charges of the four dimensional Maxwell theory. Imposing appropriate asymptotic falloff conditions, we compute the electric and magnetic soft charges and their algebra both at spatial and at null infinity. While the commutator of two electric or two magnetic soft charges vanish, the electric and magnetic soft charges satisfy a complex $U(1)$ current algebra. This current algebra through Sugawara construction yields two $U(1)$ Kac-Moody algebras. We repeat the charge analysis in the electric-magnetic duality-symmetric Maxwell theory and construct the duality-symmetric phase space where the electric and magnetic soft charges generate the respective boundary gauge transformations.  We show that the generator of the electric-magnetic duality and the electric and magnetic soft charges form infinite copies of $iso(2)$ algebra. Moreover, we study the algebra of charges associated with the global Poincar\'e symmetry of the background Minkowski spacetime and the soft charges. We discuss physical meaning and implication of our charges and their algebra.}

\maketitle

\textwidth 17cm
\evensidemargin -0.2cm
\oddsidemargin 2.2cm
\section{Introduction}
{It is by now a well known fact that gauge symmetries are not mere redundancies of description: while gauge redundancies are usually fixed by gauge fixing, a subset of the gauge group, consisting of the Large Gauge Transformations (LGT) survive the gauge fixing and act non-trivially at boundaries of the system. Indeed LGTs form an extension of global symmetries changing the state of the system. While local gauge invariant observables are unable to measure LGTs, they may be detected by nonlocal observables like the  memory effect. The role of LGTs in describing and understanding the low energy (infrared) dynamics of gauge theories has received an immense attention in recent years (e.g. see \cite{Strominger:2017zoo} and references therein). }

In this paper we focus on the four dimensional Maxwell $U(1)$ gauge theory or the QED and its LGT. The global $U(1)$ gauge transformation, the gauge transformation which approaches a non-zero constant at infinity, is the simplest LGT whose corresponding Noether charge is the usual total electric charge.
The LGT in Maxwell theory are generalization of this global transformations whose asymptotic value is determined by a scalar function on the {celestial sphere at the boundary of the space.} 

Using the standard methods,  the covariant phase space method 
\cite{Lee:1990nz,Ashtekar:1990gc, Ashtekar:1987tt}, or the Hamiltonian formulation \cite{Henneaux:1985kr, Henneaux:1992ig, Brown:1986nw, Henneaux:2018gfi, Henneaux:2018hdj}, one can associate conserved surface charges to the LGT. These surface charges, upon the equations of motion, decompose into a   ``soft'' and a  ``hard'' part \cite{Kapec:2015ena} (the latter is an integral over the external electric currents and include the usual electric charge). The soft charges  depend on the scalar LGT function on the celestial sphere and on the other hand are functions over the phase space. One can hence compute the Poisson bracket and the algebra of these charges. 
The states in the usual ``physical'' Hilbert space of QED which are specified by their usual wave-vector and polarization are now to be viewed as infinitely degenerate by the addition of these soft charges. In other words, a physical asymptotic state in a theory may have a soft-dressing. While not appearing in scattering amplitude of usual hard states, the soft-dressing may have other observable effects, e.g. as Aharnov-Bohm phase in QED or in electromagnetic \cite{Bieri:2011zb, Bieri:2013hqa, Susskind:2015hpa, Pasterski:2015zua, Hamada:2017uot, Hamada:2018vrw,Hirai:2018ijc} or gravitational memory effect \cite{Flanagan:2014kfa, Bieri:2015yia, Tolish:2014oda, Pate:2017fgt, Giddings:2018umg}. Moreover, as first noted by Faddeev and Kulish \cite{Kulish:1970ut} and reemphasized recently (see e.g. \cite{Gabai:2016kuf, Herdegen:2016bio} for a nice overview and summary of this issue), a specific soft-dressing for the charged states or the vacuum may be needed to satisfactorily address the IR issues in gauge theory or gravity.

Maxwell theory  enjoys Electric-Magnetic Duality (EMD). In the simplest version this duality is a $Z_2$ which exchanges electric and magnetic fields while it can be promoted to a $U(1)$ symmetry, continuously rotating the electric and magnetic fields into each other. Moreover, by the addition of the $\theta$-term this $U(1)$ may be extended to $SL(2,R)$ at the classical level. The symmetry between the electric and magnetic descriptions is, however, broken in QED when we introduce the electric charge into the system where we choose to work with electric degrees of freedom. Nonetheless, one may also introduce magnetic charge and currents to maintain the symmetry. The EMD is known to extend to non-Abelian gauge theories and in particular the $SL(2,Z)$ part of it,  remains an exact quantum symmetry in the context of supersymmetric gauge theories \cite{Montonen:1977sn, Seiberg:1994rs}.

In this work we revisit the question of soft charges in the context of electric-magnetic duality. Inspired by this duality, the magnetic dual of soft charges were proposed in \cite{Strominger:2015bla} and the corrections to soft theorems in the presence of magnetic charges was derived from the conservation of magnetic soft charges. However, it was shown later \cite{Campiglia:2016hvg} that the magnetic soft charges have a key role even in the absence of magnetic sources. Indeed Weinberg's soft theorem implies the existence and conservation of both electric and magnetic soft charges. This raises the question what is the nature of magnetic soft charges. In Maxwell theory the electric soft charges appear as the Noether charges associated to a set of LGT of the theory while the magnetic soft charges have no LGT counterpart. Moreover, the boundary conditions usually used in construction of radiative phase space of Maxwell theory excludes magnetic sources and accordingly magnetic soft charges. In section \ref{DS section}, we consider the duality symmetric formulation of Maxwell theory \cite{Zwanziger:1968rs, Zwanziger:1970hk} which gives an answer to the above two questions at the same time. While the duality symmetric theory is on-shell equivalent to the Maxwell in the bulk, there is an extra boundary gauge symmetry whose conserved  charges are exactly the magnetic soft charges. We therefore have two sets of electric and magnetic soft charges and the associated electric and magnetic LGTs. The duality symmetric formulation enables us to construct the duality symmetric phase space and allows us to put the electric and magnetic soft charges at the same footing. 
 
Our main, perhaps surprising, result is while the electric soft charges (and similarly magnetic soft charges) commute among themselves and form an Abelian algebra, electric and magnetic soft charges associated with certain electric and magnetic LGT do not commute with each other. To ensure that our results are not artifacts of  the way we perform the analysis, we make the calculations in some different ways: (1) we compute the charges both at null and spatial infinities; (2)  we use the duality invariant  Maxwell theory to compute the charges and their algebra. All these of course yield to the same result.

The rest of this paper is organized as follows. In section \ref{sec:2}, we compute the electric and magnetic soft charges in the Lorenz gauge by imposing appropriate falloff behavior at spatial infinity. We show that these soft charges could be viewed as Hamiltonian generators on a phase space and using this we compute the algebra of electric and magnetic soft charges. We show if we allow LGT which are nonregular at the celestial sphere, electric and magnetic charges do not commute. We note that these singular gauge transformations are inevitable if we have charged particles going through the null infinity \cite{Cardona:2015woa}. In section \ref{Null-section}, we repeat the analysis of section \ref{sec:2} but compute the charges at asymptotic future null infinity. The analysis of this section reconfirms the same charge algebra. In section \ref{DS section},  
we consider an extension of the Maxwell theory which is invariant under the electric-magnetic duality and compute the soft charges in this theory. In this case the electric and magnetic soft charges appear at the same footing. In particular, we discuss the charge associated with the $U(1)$ global symmetry rotating electric and magnetic fields into each other, the duality charge. We work out the algebra of this ``duality charge'' and the soft charges. Moreover, we analyzed conserved (Noether) charges associated with Poincar\'e symmetry and study the algebra of soft charges, the Poinar\'e charge and the duality charge and show that the spin (angular momentum) charge is different than the duality charge.  Section \ref{discussion:sec} is devoted to a summary, discussion and physical implication of our results and the outlook. In appendix \ref{App:A}, we have gathered some technical details of the charge integrals. In appendix \ref{App:B}, we discuss complexified Maxwell theory as a variant formulation the duality symmetric theory.

\paragraph{Notations and conventions.}  We will be working with a gauge field theory, with dynamical field one-form $A$, and with the field strength two-form $F=dA$. We decompose this two-form into a spatial two-form, the magnetic field $B$, and a spatial vector, the electric field $E$.  

Field configurations on a constant time slice $t$, $A(x;t)$, as we discuss, parameterize the covariant phase space (a la Wald \cite{Lee:1990nz}).  Tangent space to this phase space can be spanned by generic field variations. In our conventions, we denote the variations which are one-forms on the phase space as $\df A$. In general, hence, we are dealing with forms on spacetime and the phase space. By a $(p;q)$-form we mean a $p$-form in spacetime and a $q$-form on the phase space. Exterior derivative on the spacetime and phase space will be respectively denoted by $d$ and $\df$. So, given a $(p;q)$-form $X$, $d X$ is a $(p+1;q)$-form and $\df X$ a $(p;q+1)$-form. The two spacetime and phase space exterior derivatives commute with each other, $d\df X=\df dX$. 

Besides the forms, we also have vectors on the phase space which we denote by $\delta$. In particular, we denote the vector associated with the function $f$  by  $\delta_f$. The interior product between forms and vectors on the phase space  will be denoted by ${\di}$, e.g. given the $(p;q)$-form $X$, $\mathbbmss{i}_{\delta_f}X$ is a $(p;q-1)$-form.\footnote{In the notation more common in the literature of this field, the exterior phase space derivative is denoted by $\delta$, rather than $\df$. However, this usual notation does not  distinguish between the vector and forms on the phase space.} We also define Lie derivative on the phase space along a generic vector $\delta$ and denote by $\mathbbmss{L}_\delta$. Given a function on the phase space (i.e. a $(p;0)$-form) $\phi$,
\be
\mathbbmss{L}_\delta \phi=\delta\phi,
\ee
is nothing but the usual variation of $\phi$. Phase space Lie derivative on generic $(p;q)$-form $X$ can then be defined through  the Cartan identity:
\be
\mathbbmss{L}_\delta X=\df(\di_\delta X)+\di_\delta\df X, 
\ee
and one may show that $\mathbbmss{L}_\delta \df X=\df(\mathbbmss{L}_\delta X)$ for any $X$.

Hodge-star operation denoted by $\ast$, is defined only on  spacetime or just the spatial part; the latter will of course be manifest from the context. Therefore in $d$ space(time) dimensions and for a generic $(p;q)$-form $X$, $\ast X$ is a $(d-p;q)$-form. As it is clear, $\df$ commutes with Hodge-star operation $\ast$, $\ast \df X=\df \ast X$.

We use the same notation $\wedge$-product for both  spacetime
and phase space forms. That is,  for a $(p;q)$-form $X$ and  a $(r;s)$-form $Y$,
\be
X \wedge Y =(-1)^{pr}(-1)^{qs}\ Y \wedge X,
\ee
and
\be
d(X\wedge Y)=dX\wedge Y+(-1)^p X\wedge dY,\qquad \df(X\wedge Y)=\df X\wedge Y+(-1)^q X\wedge \df Y.
\ee

We will introduce the rest of notations used in the main text, when they appear.

\textbf{Note added.} Soon after our paper, the reference \cite{Freidel:2018fsk} appeared on arXiv. While the approaches are different, interestingly the main results agree.
\section{Maxwell soft charges at spatial infinity}\label{sec:2}
Consider the Maxwell theory in $d$ dimensional spacetime described by the gauge field one-form $A=A_\mu dx^\mu$ and the field strength two-form $F=dA$, governed by the action
\begin{align}\label{Lagrangian}
    S=-\frac14\int  F_{\mu\nu}F^{\mu\nu}= -\frac{1}{2}\int F \wedge \ast F\,.
\end{align}
To compute the charges one may use the covariant phase space approach \cite{Lee:1990nz,Ashtekar:1990gc, Ashtekar:1987tt}. To this end we  study variation of the Lagrangian with respect to generic field variations, yielding the equations of motions and a total derivative term: 
\begin{eqnarray}\label{Lagrangian variation}
\df {\cal L} =-d \ast F \wedge \df A  +d \Theta,
\end{eqnarray}
where $\Theta=-\ast F\wedge \df A$ is the (pre)symplectic potential density. In the language of forms, the Lagrangian is a $(d;0)$-form, $\df L$ a $(d;1)$-form  and $\Theta$ a $(d-1;1)$-form.
While the equations of motion determine dynamics of the system, the second term induces the symplectic structure of the covariant phase space, see  \cite{Seraj:2016cym,Compere:2018aar} for reviews. From this, one defines the presymplectic current as a $(d-1;2)$-form
\begin{eqnarray}
\omega=\df \Theta=-\ast \df F\wedge \df A\,.
\end{eqnarray}
 The presymplectic structure $\bOmega$ of the theory is the integration of presymplectic current over a hypersurface $\Sigma$,
\begin{eqnarray}\label{Symplectic}
\bOmega=\int_{\Sigma}\omega=-\int_{\Sigma}\ast \df F\wedge \df A\,.
\end{eqnarray}
To guarantee that the phase space contains all degrees of freedom,  $\Sigma$ must be a Cauchy surface.  The above is called the presympectic form as it has degeneracies: $\bOmega$ vanishes for field variations without support on $\Sigma$. The phase space with a nondegenerate symplectic form is then obtained by a symplectic quotient by the degeneracies \cite{Lee:1990nz,Woodhouse:1980pa}. However, this simply means that one should restrict attention to those configurations and associated variations that have support on the Cauchy surface $\Sigma$, i.e they do not vanish on $\Sigma$. Hereafter, we will only consider such configurations over which the $(0;2)$-form $\bOmega$ is the symplectic form.

\subsection{Electric and magnetic soft charges}

\paragraph{Electric (Noether) soft charges.} A gauge transformation $A \rightarrow A+df$ induces a vector field $\delta_f$ over the space of fields.  One can then define the Hamiltonian generator $\bQ_f^E$ associated with this gauge transformation,
\begin{align}
\df \bQ_f^E=-\di_{\delta_f}\bOmega\,.
\end{align}
$\df \bQ^E_f$ is a $(0;1)$-form and the charge (Hamiltonian generator) $\bQ^E_f$ exists if $\df \bQ$ is integrable, that it is an exact $(0;1)$-form. From (\ref{Symplectic}) we find,  
\begin{align}\label{electric-charge-var}
\df \bQ_f^E=\int_{\Sigma}\di_{\delta_f}\big[\ast \df F\wedge \df A\big]=
-\int_{\Sigma}\ast \df F\wedge df\,.
\end{align}
Since $f$ is constant over the space of fields, the charge variation \eqref{electric-charge-var} is integrable and one can simply integrate the above relation and write the Hamiltonian generator as 
\begin{align}\label{EHamiltonian}
\bQ_f^E=-\int_{\Sigma} \ast F\wedge df.
\end{align}
Had we computed the above on-shell, it simply reduces to the Noether electric charge associated to the gauge transformation $A\to A+df$. We shall comment on this further below.

The Hamiltonian generator (\ref{EHamiltonian}) and the symplectic structure (\ref{Symplectic})
can be written in terms of electric and magnetic fields $E$ and $B$. This can be provided by taking 
the Minkowski spacetime as $\mathcal{M}=\mathds{R}\times \Sigma_t$ through choosing
a time function $t$ and working in a coordinate such that the metric can be written as 
\be
ds^2=-dt^2+h_{ab}dx^adx^b,\qquad x^a, \ a=1,\cdots, d-1,
\ee
where
$h_{ab}$ is a Reimannian metric on $\Sigma_t$ and $x^a$ are coordinates on it. To be more explicit we will denote the Cauchy surface at constant time slice by $\Sigma_t$.  With this decomposition, we can split $F$ as,
\begin{align}\label{decomp-F}
F=B+E\wedge dt
\end{align}
where $B$ and $E$ are differential forms of ranks two and one on $\Sigma_t$, representing the magnetic and electric fields respectively. With the convention $\epsilon^{0123}=1$ for the Levi-Civita tensor, we have $\ast F=-\ast E+\ast B\wedge dt$, where the $\ast$ in the left-hand-side is a four dimensional Hodge star and the one in the right-hand-side is a three dimensional one. The conventional electric and magnetic vector fields are related to the differential  forms $E=E_a\, dx^a, B=B_{ab}\,dx^a\wedge dx^b$ as\footnote{We use the same notation for space(time) forms and vector fields, which should be understood from the location of indices. }
\begin{align}
    E^a=h^{ab}E_b,\qquad B^a=\dfrac{1}{2}\varepsilon^{abc}B_{bc}.
\end{align}
The symplectic structure $\bOmega$ and the generators in terms of this decomposition are, 
\begin{align}
\bOmega&=\int_{\Sigma_t}\ast \df E\wedge \df A=\int_{\Sigma_t}d^{d-1}x\sqrt{|h|}\ \df E^a\wedge \df A_a\label{symplectic-1}\\
\bQ_f^E&=\int_{\Sigma_t} \ast E\wedge df=\int_{\Sigma_t}d^{d-1}x\sqrt{|h|}\ E^a \partial_a f\label{QE-Hamiltonian-generator}\,.
\end{align}
where $\ast$ is Hodge dual operator and must be understood appropriately when acts on  forms on $\mathcal{M}$ like $F$ or on spatial forms on $\Sigma_t$ like $E$ and $B$. Note that $\bQ_f^E$ is written completely in terms of electric field and this justifies the index $E$.

\begin{figure}[t]
    \captionsetup{width=.8\linewidth}
    \centering
    \includegraphics[scale=.7]{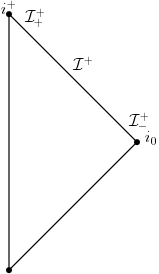}
    \caption{Penrose diagram of 4d flat space. ${\cal I}^+_\pm$ denote the past and future boundaries of the null infinity ${\cal I}^+$. These boundaries are essentially future time like infinity $i^+$ and the spatial infinity $i^0$.}
    \label{Penrose-diagram}
\end{figure}

\paragraph{Notation:} So far we did not restrict our fields or their variations to any field equation. Since in our analysis we will need to also impose equations of motion we introduce the following notation. 
The off-shell quantities will be denoted by boldface symbols, while their on-shell value will be denoted with the same notation but not in bold. For example the Hamiltonian generators $\bQ_f^E$ denotes the generator of gauge transformation $\delta_f$ in the bulk, while its on-shell value $Q_f^E$ is the corresponding charge. Similarly, $\bOmega$ is the off-shell symplectic form, while $\Omega$ refers to the projection to the space of solutions. Moreover, $\approx$ means equality on-shell. For example,
\begin{align}\label{ECharge}
\bQ_f^E \approx Q^E_f= -\oint_{\partial\Sigma} f \ast F=\oint_{\partial\Sigma_t} f \ast E=\oint_{\partial\Sigma_t} d^{d-2}x\sqrt{|h|}\ f  E^a n_a\,,
\end{align}
where  $\partial\Sigma_t$ is the boundary of the Cauchy surface $\Sigma_t$ and $n_a$ is the vector normal to the $\partial\Sigma_t$. In our analysis, as is usual, we decompose the spatial metric $h$ (metric on $\Sigma_t$) as $h=n\otimes n+{\cal G}$ where $n=dr$ and $r$ is a radial coordinates ($0\leq r<\infty$) and ${\cal G}$ is conformal to the metric on the celestial sphere. Here $r \rightarrow \infty$ corresponds to $\partial \Sigma_t$ which is the boundary of the Cauchy surfaces and in our analysis is the spatial infinity $i^0$, \emph{cf.} figure \ref{Penrose-diagram}.

Note that whether $Q^E_f$ is zero or not, depends on the behavior of the associated gauge symmetry $f$ at the boundary.
For the LGT which are non-zero at  $\partial\Sigma_t$,  $Q^E_f$ is generically non-zero. These transformations hence label the soft charges of the phase space. Finally we note that hereafter we restrict ourselves to $d=4$ and to Maxwell theory on four dimensional flat spacetime.

\paragraph{Magnetic soft charges.} Motivated by the electric-magnetic duality, one can define magnetic dual of the infinite electric conserved charges $Q^E_f$'s in (\ref{ECharge}). {To this end we note that $J^B_g=F \wedge dg$ is a conserved quantity, $dJ^B_g=0$, as a result of the Bianchi identity $dF=0$. Therefore, one may define
the conserved magnetic charge as
\be
\int_{\Sigma} J^B_g=\int_{\Sigma}F\wedge dg=\oint_{\partial\Sigma} gF=\oint_{\partial\Sigma_t} d^2x\sqrt{|h|}\ n_a  B^a \,g,
\ee
where $g$ is a function on the spacetime. On the other hand, the Bianchi identity $dF=0$ can be locally solved in terms of the gauge potential $A$, $F=dA$. Inserting this into the above integral we learn that the above conserved charge will have an integral on celestial  sphere and a contour integral (\emph{cf.} \eqref{bQB-vs-QB}). For $g=1$ case the above integral is nothing but the total magnetic charge of the system.
With the discussion above, one may then propose a Hamiltonian generator for magnetic charges $\bQ_g^B$ which is a function of the gauge potential $A$, as
\begin{align}\label{BCharge}
\bQ^B_g\equiv \left\{\begin{array}{cc} \oint_{\partial\Sigma} A\wedge dg, &\qquad {g\neq 1}\\ & \\ \oint_{\partial\Sigma} B= \oint_c A, & \qquad g=1,\end{array}\right.
\end{align}
where $\partial\Sigma$ denotes the boundary of Cauchy surface, the celestial sphere, and $c$ denotes a contour on the sphere which encircles all the singularities. The $g=1$ expression, as pointed out, measures the total magnetic charge of the system, see figure \ref{CSphere}. A system with non-zero magnetic charge can be modeled by a usual Dirac string \cite{Jackson:1998nia}. The Dirac string is described by a gauge potential which has a jump (is not single-valued) and $\oint_c A$ in fact measures this jump, aka the magnetic charge. In this sense the $g=1$ measures the magnetic ``hard charge''. The $g\neq 1$ expression, however, measure the  ``magnetic soft charges''. As will become clear below, the non-zero contributions to $Q^B_g$ \eqref{BCharge}  can be compensated by singular-large gauge transformations, i.e. for $A=d\lambda$ where $\lambda$ is non-vanishing on the celestial sphere and has localized singularities. The latter may be viewed as points the Dirac strings of the 3d bulk hits the celestial sphere. Note also that, in contrast to the electric case, the magnetic current is not a Noether current and does not stem from a (gauge) symmetry of the usual Maxwell theory.\footnote{As we will discuss in section \ref{DS section}, in the dual symmetric description of the Maxwell theory $\bQ^B_g$ is also promoted to a Noether symmetry.} We shall discuss this point further in the last section, after discussing the dual symmetric Maxwell theory in section \ref{DS section}.

\paragraph{Algebra of charges.}  From the symplectic structure \eqref{symplectic-1},
we have the following equal time $t=const$ Poisson brackets,
\begin{align}\label{Maxwell Brackets spatial}
\{E_a(x),A_b(x')\}_t=\frac{h_{ab}}{\sqrt{|h|}}\delta^3(x-x')\,,\,\,\{A_a(x),A_b(x')\}_t=0\,,\,\, \{E_a(x),E_b(x')\}_t=0.
\end{align}
With the above we can compute the Poisson bracket of $\bQ^E$ and the gauge field $A_a$,
\be
\{\bQ^E_f, A_a(x)\}=\partial_a f, \qquad x\in \Sigma_t.
\ee
The above means that $\bQ^E_f$ is generator of gauge transformations on $\Sigma_t$ (this is in accord with the name Hamiltonian generator). We note that $\bQ^B_g$ is not generator of a gauge transformation in the bulk. 
Nonetheless, one may check that
\begin{align}\label{QB-E}
    \{\bQ^B_g,\vec{E}(x)\}=
        \hat{r}\times \nabla g(x) \ \delta(r-R) 
\end{align}
where $r$ is the radial coordinate transverse to the boundary of the Cauchy surface $\partial\Sigma$ and  the asymptotic boundary is located at $r=R$. 
If $\vec{E}$ at the boundary is of the form $\nabla\times C$ then $\bQ^B_g$ generates gauge transformations on $C$. 
We shall return to this in more detail in section \ref{DS section}. 
One can also compute Poisson bracket of electric and magnetic Hamiltonian generators,
\begin{align}
  \nonumber  \{\bQ_{f}^E,\bQ_{g}^B \}_t&=\int_{\partial\Sigma_t}  d^2x' \int_{\Sigma_t} d^3x\sqrt{|h|}\  \epsilon^{rab}\{E^c(x),A_a(x')\}\partial_c f(x) \partial'_b g(x')\\
  \nonumber  &=\oint_{\partial\Sigma_t} d^2x\  \epsilon^{ab}\partial_af\partial_b g\\
    &=\oint_{\partial\Sigma_t} df\wedge dg\,,\label{central extension}
\end{align}
for generic $g$ and $\{\bQ_{f}^E,\bQ_{g=1}^B \}_t=-\oint_c df$. This latter integral is nonzero only if $f$ has a cut; it is not single-valued as we go round the contour $c$.
In a similar way one can compute the algebra of electric and magnetic Hamiltonian generators, 
\begin{align}\label{Poisson bracket-spatial}
\{\bQ_{f_1}^E,\bQ_{f_2}^E \}_t=0\,, \qquad \{\bQ_{g_1}^B,\bQ_{g_2}^B \}_t=0, \qquad \{\bQ_{f}^E,\bQ_{g}^B \}_t=\oint_{\partial\Sigma_t} df\wedge dg=-\oint_c gdf\,.
\end{align}

To understand the above result better, we recall  that while the magnetic soft charge acts trivially in the bulk, it generates an electric field on (and tangent to) the boundary which can be written as the gradient of a boundary potential $g\big{|}_{\partial\Sigma}$. This anomalous boundary field leads to the central extension \eqref{central extension} in the algebra of electric and magnetic soft charges if one of $f$ or $g$ is nonsmooth at the boundary. A similar argument may be repeated considering $\{\bQ^E_f,\vec{B}(x)\}$. In this case, $Q^E_f$ generates a gauge transformation $A\to A+df$ and therefore $\vec{B}\to \vec{B}+\nabla\times \nabla f$ which is trivial when $f$ is a smooth function. However, we will see in the next section that the gauge parameters of interest are holomorphic functions with poles at some points on the boundary. In that case, one can show that the singular gauge transformation generates a set of Dirac strings with both ends at the boundary. Each endpoint resembles a magnetic (multi)pole at the boundary. We will discuss this point as we go along and in particular in section \ref{discussion:sec}.

\subsection{On-shell covariant phase space and boundary symmetries}\label{sec:on-shell-maxwell}

In previous section we introduced the covariant phase space built on the configuration space of \textit{histories} $A(x;t)$ and defined through the symplectic structure \eqref{Symplectic} or \eqref{symplectic-1}.\footnote{The covariant phase space may be compared with the usual phase space appearing in the Hamiltonian formulation which is built over the space of instantaneous configurations $A(x)$ and their canonical conjugates $\Pi(x)$.} The set of solutions to the field equations may be (heuristically) considered as a submanifold in the configuration space and field variations $\df A(x;t)$ are one-forms on this phase space. Once pulled-back on the solutions submanifold where we impose equations of motion on the field and  the linearized field equations on field variations, however, the sympelctic structure is not invertible \cite{Lee:1990nz}.  To see this consider the contraction of the symplectic form with a gauge transformation $\lambda$ with compact support on $\Sigma_t$, i.e. $\lambda=0$ at the boundary $\partial\Sigma_t$. Then,\footnote{Recall that according to our notation $\Omega$ refers to the pull-back of the symplectic form $\bOmega$ onto the space of solutions. See the remark above equation \eqref{ECharge}. }
\begin{align}
\di_{\delta \lambda}\bOmega=\int_{\Sigma_t}  \df\ast F \wedge d\lambda= \oint_{\partial\Sigma_t}\lambda\,\df\ast F- \int_{\Sigma_t} \lambda \, d (\df\ast F).
\end{align}
While nonvanishing in general, $\di_{\delta \lambda}\bOmega$ vanishes on the solution submanifold.  In other words, local gauge transformations are null directions of the structure on the solution submanifold. One can consider the solution space as a fiber bundle whose fibers are generated by local gauge transformations. The symplectic quotient over degeneracies then corresponds to working with equivalence classes (base manifold), or equivalently choosing a section of the bundle, i.e. gauge fixing. This procedure, however, leaves us with \emph{large (boundary) gauge transformations} considered as physical symmetries of the phase space, as for $\lambda$'s with support on the boundary  $\di_{\delta \lambda}\Omega=-\oint_{\partial\Sigma}\lambda\df\ast F\neq0$. This is the on-shell  covariant phase space \cite{Ashtekar:1990gc, Lee:1990nz}.

 To see the above construction explicitly, let's write $A=\hat A+d\psi$ where $\hat A$ is the divergence-free part of the gauge field. Gauge transformation corresponds to $\hat A\to \hat A, \psi\to \psi+\lambda$. The symplectic structure is then given by 
\begin{align}\label{symplectic-Maxwell-onshell}
   \hspace*{-5mm}  \Omega= -\int_{\Sigma_t}\ast \df F\wedge \df \hat A-\oint_{\partial\Sigma_t} \df \ast F\wedge \df \psi=\int_{\Sigma_t} d^3x\sqrt{h} \;\df \hat E^i\wedge \df \hat A_i + \oint_{\partial\Sigma_t} d^2x\sqrt{\gamma} \;\df ( E.n) \wedge \df \psi\,.
\end{align}
where in the last equality $\hat E^i$, the transverse part of the electric field, appears as we have already imposed the constraint equation $\nabla\cdot E=0$. The symplectic reduction is manifest in the fact that only the boundary value of $\psi$ matters in the symplectic form. Moreover, we observe that the final symplectic structure breaks into the bulk radiative phase space of transverse photons and the boundary phase space involving the boundary field $\psi$  with its canonical pair being the normal component of the electric field. We will see in section \ref{DS section} that in the dual symmetric version of Maxwell theory, an extra magnetic boundary mode will naturally appear.

\subsection{Asymptotic symmetries and their algebra}
To make the general construction of the previous sections explicit we need to specify the configuration space under consideration and the set of LGT's.  This is usually done by a suitable choice of boundary conditions and possibly a gauge fixing. 
We work in the coordinate system $(t,r,z,\bar{z})$ in which the Minkowski metric $ds^2=-d\mathrm{x}_0^2+d\mathrm{x}_1^2+d\mathrm{x}_2^2+d\mathrm{x}_3^2$ takes the form\footnote{The map between these coordinates is, $t=\mathrm{x}_0, \  r^2=\mathrm{x}_1^2+\mathrm{x}_2^2+\mathrm{x}_3^2,\ 
z=\frac{\mathrm{x}_1+i\mathrm{x}_2}{r+\mathrm{x}_3}.$}
\begin{align}
ds^2=-dt^2+dr^2+2r^2 \gamma_{z\bar z} dz d\bz,    
\end{align}
where  $z,\bar z$ are the stereographic coordinates on the celestial sphere and $\gamma_{z\bar z}=\frac{2}{(1+z\bar{z})^2}$. By this, we have essentially removed a point (the south pole) from the sphere. An integral over the sphere then maps to an integral over the complex plane with the boundary (the south pole of the sphere) at $z\to \infty$. 

We now propose the following boundary conditions on components of electric and magnetic fields near spatial infinity $r\to\infty$,
\begin{align}\label{BC1}
&E_{z,\bz} \sim O(\frac{1}{r}),\qquad E_r \sim O(\frac{1}{r^2}),\\ 
&B_{z,\bz} \sim O(\frac{1}{r}),\qquad B_r \sim O(\frac{1}{r^2}),\label{BC3}
\end{align} 
These falloff conditions (\ref{BC1}) and \eqref{BC3} ensure that the electric and magnetic charges are finite in the bulk. With these boundary conditions 
the symplectic flux is vanishing at the spatial infinity, yielding the conservation of electric charges. %
These boundary conditions can be derived by the following boundary conditions at spatial infinity $i^0$ on the components of gauge field $A$ and its time derivative,
\begin{align}\label{BC-gauge-field-A}
A_t \sim O(\frac{1}{r})\,, \qquad A_r \sim O(\frac{1}{r})\,,  \qquad A_{z,\bz} \sim O(1)\,, \qquad \partial_t A_r\sim O(\frac{1}{r^2}),\qquad \partial_t A_{z,\bz}\sim O(\frac{1}{r})\,.
\end{align} 
The above can be more explicitly expressed as
\begin{align}\label{BC-gauge field}
A_t=\sum_{n=1} \frac{A_t^{(n)}(t,z,\bz)}{r^n}\,, \quad A_r=\sum_{n=1} \frac{A_r^{(n)}(t,z,\bz)}{r^n}\,, \quad  A_{z,\bz}=\sum_{n=0} \frac{A_{z,\bz}^{(n)}(t,z,\bz)}{r^n}\,, \qquad r\rightarrow \infty
\end{align} 
where $A_r^{(1)}$ and $A_{z,\bz}^{(0)}$ are independent of $t$.
To proceed we fix the Lorenz gauge, $\nabla_{\mu} A^{\mu}=0.$  This leaves us with the set of residual gauge symmetries $\lambda$ satisfying,
\begin{align}
\square \lambda=-\partial_t^2 \lambda+\partial_r^2 \lambda+\frac{2}{r}\partial_r \lambda+\frac{2}{r^2\gamma_{z\bz}}(\partial_z \partial_{\bz} \lambda)=0 \,.
\end{align} 
Expanding $\lambda$ near the boundary $\partial \Sigma_t$ as 
\begin{align}
\lambda(t,r,z,\bz)=\sum_{n=0} \frac{\lambda^{(n)}(t,z,\bz)}{r^n}\,, \qquad r\rightarrow \infty,
\end{align} 
up to third order in expansion, we have
\begin{align}
\partial_t^2 \lambda^{(0)}=0,\qquad &\partial_t^2 \lambda^{(1)}=0,\\
\partial_t^2 \lambda^{(2)}-\frac{2}{\gamma_{z\bz}}(\partial_z \partial_{\bz} \lambda^{(0)})=0,\qquad &
\partial_t^2 \lambda^{(3)}-\frac{2}{\gamma_{z\bz}}(\partial_z \partial_{\bz} \lambda^{(1)})=0.\label{residual-sphere}
\end{align} 
The behavior of $A_t$ in \eqref{BC-gauge-field-A} implies that $\partial_t \lambda^{(0)}=0$ and hence
\begin{align}\label{f2}
     \lambda^{(2)}=t^2\frac{\partial_z \partial_{\bz} \lambda^{(0)}}{\gamma_{z\bz}}+t\alpha(z,\bz)+\beta(z,\bz).
\end{align}

Moreover, we require the energy on a constant time slice $t$ of the configurations to remain finite even in the limit $|t|\to \infty$. This is satisfied if 
\begin{align}\label{asymptotic condition-spatial}
    A_a(t,x)\sim \mathcal{O}(t^0),\qquad t\to \pm \infty.
\end{align}
This further constrains the symmetries. An LGT respecting \eqref{asymptotic condition-spatial} must satisfy $\partial_a\lambda=\mathcal{O}(t^0)$ implying $\nabla_{S}^2 \lambda^{(0)}=const$, where $\nabla_{S}^2$ is the Laplacian on the (unit) sphere. The eigenvalues of this Laplace operator are either zero, corresponding to eigenfunctions solving $\partial_z\partial_{\bz}\lambda^{(0)}=0$ or negative corresponding to spherical harmonics studied in \cite{Seraj:2016jxi}. In this paper, we are interested in the former, so we take
\begin{align}\label{LGTs}
\partial_t \lambda^{(0)}=0\,, \qquad \partial_z \partial_{\bz} \lambda^{(0)}=0.
\end{align}

Assuming smoothness, the solutions are given by holomorphic and antiholomorphic functions on the sphere. However, the only such function on the sphere is a constant function. Therefore, we relax the smoothness condition and allow $\lambda$ to have singularities at finite number of points on the sphere.  Locally, a complete basis is 
\begin{align}\label{complex-basis-f}
f_P=\ln z,\qquad  f_n=z^n,\quad n\in \mathbb{Z},
\end{align}
which have typically  poles at the north pole $z=0$. Note that our coordinate system $z,\bz$ does not cover the south pole which is the boundary of our chart on the sphere. 
Let us denote the charges associated with the gauge parameters $f_n, f_P$ by $\bQ_n, \mathbf{P}$ respectively. Using \eqref{Poisson bracket-spatial} and the formulas in the appendix, we find that 
\begin{align}\label{algebra-spatial}
\{\bQ_{n}^E,\bQ_{m}^B\}=2\pi i m \delta_{m+n,0},\qquad 
\{\bQ^B_n,\mathbf{P}^E\}=2\pi i\delta_{n,0},\qquad \{\bQ^E_n,\mathbf{P}^B\}=2\pi i\delta_{n,0}.
\end{align}
Note that  $\mathbf{P}^E$ is paired with the magnetic charge generator $\bQ^B_0$. This means that a singular gauge transformation with a logarithmic gauge parameter produces a magnetic charge. This is indeed the manifestation of the Dirac monopole construction. For a similar discussion see \cite{Cardona:2015woa}. {More specifically, we note that a logarithmic term in the gauge transformation means it is not single-valued as we encircle $z=0$; the amount of jump is proportional to the magnetic charge \cite{Jackson:1998nia}.}
Likewise we note that 
\begin{align}
    \partial\bar\partial z^{-n}=\dfrac{2\pi i}{(-1)^n n!}\partial^n \delta^2(z), \qquad n\geq 1.
\end{align}
This implies that a gauge transformation of the form $f=z^{-n}\  (n\geq 1)$ generates a source of the form $\partial^n \delta^2(z)$ at the origin. This is nothing but a multipole charge of order $n$, i.e. $n=1$ generates a dipole moment, $n=2$ a quadrupole, etc. A monopole charge, on the other hand is generated by the logarithmic gauge parameter, since $\partial\bar\partial\ln z=2\pi i\delta^2(z)$.

\paragraph{Conservation of charges.}
It is straightforward to see the pull-back of the symplectic density \eqref{symplectic-1} at the boundary (i.e. at constant $r=R\gg 1$ surface) is,
\begin{align}
\bOmega_{R}&=\int_{R} \omega= \int_{R}dt d^2z \big[\df F_{rz}\wedge \df A_{\bz}+\df F_{r\bz}\wedge \df A_{z}+
r^2 \gamma_{z\bar z} \df F_{tr} \wedge \df A_{t}\big]\nonumber \\
&=\int_{R} dt d^2z\,  \big[ i (\df B_{z}\wedge \df A_{\bz}-\df B_{\bz}\wedge \df A_{z}) -
r^2 \gamma_{z\bar z} \df E_{r} \wedge \df A_{t}\big]\,,
\end{align}  
where $d^2z \equiv -i dz \wedge d \bz$ and in our conventions  $\tilde{\epsilon}^{trz \bz}=i$ where $\tilde{\epsilon}^{\alpha \beta \mu \nu }$ is the Levi-Civita symbol.\footnote{From this convention choice it follows  that $\tilde{\epsilon}_{trz \bz}=i$.}.  Recalling our boundary conditions, this leads to,
\begin{align}
\bOmega_{R} \sim O(\frac{1}{R}).
\end{align} 
The electric charges are hence conserved. In the above analysis conservation of magnetic charge is evident as the magnetic field is absent in the symplectic structure.

An alternative argument for charge conservation is as follows.
Recall that
\begin{align}
J^E_f=\ast F\wedge df\,, \qquad J^B_g= F\wedge dg,
\end{align}
are conserved current of electric and magnetic charges,
\begin{align}
dJ^E_f=0\,, \qquad dJ^B_g=0\,.
\end{align}
The flux of electric and magnetic charges is then given by
\begin{align}
\mathcal{F}_E=\lim_{R\rightarrow \infty} \int_{R} J^E_f=\lim_{R\rightarrow \infty}\int_{R} dt d^2z\, \big[i (B_{z} \partial_{\bz}f-B_{\bz}\ \partial_{z}f)-r^2 \gamma_{z\bar z} E_{r} \partial_t f \big]\sim O(\frac{1}{R})\to 0,
\end{align}
and 
\begin{align}
\mathcal{F}_B=\lim_{R\rightarrow \infty} \int_{R} J^B_g=\lim_{R\rightarrow \infty}\int_{R} dt d^2z\, \big[i (E_{z} \partial_{\bz}g-E_{\bz}\ \partial_{z}f)+r^2 \gamma_{z\bar z} B_{r} \partial_t g \big]\sim O(\frac{1}{R})\to 0,
\end{align}
where we used our falloff behavior \eqref{BC1} and \eqref{BC3}.
So, the charges are really conserved.

\section{Maxwell soft charges at null infinity}\label{Null-section}

In this section we repeat studying the electric and magnetic soft charges computed at null infinity, most conveniently analyzed in the $(u,r,z,\bar z )$ coordinate system\footnote{This is related to the Cartesian coordinates as, $r^2=\mathrm{x}_1^2+\mathrm{x}_2^2+\mathrm{x}_3^2, \, u=\mathrm{x}_0-r,\, 
z=\frac{\mathrm{x}_1+i\mathrm{x}_2}{r+\mathrm{x}_3}.$}
\be
ds^2=- du^2-2dudr+ r^2\gamma_{z\bar z} dzd\bar z,
\ee
in which 
\be
A^u=-A_r,\quad A^r=A_r-A_u,\qquad A^z=\frac{1}{r^2\gamma_{z\bar z}}A_{\bar z},\quad A^{\bar z}=\frac{1}{r^2\gamma_{z\bar z}}A_{z}.
\ee
In this coordinate system the future null infinity ${\cal I}^+$ is given by $r\to\infty$. The coordinate $u$ parametrizes the null direction on ${\cal I}^+$ and its boundaries at $u\to\pm\infty$ are denoted by ${\cal I}^+_\pm$ as depicted in figure \ref{Penrose-diagram}. 

Our convention is $\tilde{\epsilon}^{urz \bz}=i$, which leads to $\tilde{\epsilon}_{urz \bz}=i$, so it follows that,
\begin{align}\label{starF}
(\ast F)_{uz}=i(F_{uz}-F_{rz}),\qquad (\ast F)_{u\bar z}=-i(F_{u\bar z}-F_{r\bar z}),\qquad   (\ast F)_{z\bar z }=-ir^2 \gamma_{z\bar z}F_{ur}.
\end{align}
 
We start with fixing the Lorenz gauge, 
\begin{align}\label{Lorenz-gauge}
-\partial_u A_r-\partial_r A_u+\partial_r A_r+\dfrac{2}{r}(A_r-A_u)+\frac{1}{r^2\gamma_{z\bar z}}(\partial A_{\bar z}+\bar\partial A_z)=0,
\end{align}
and imposing the asymptotic falloff conditions \cite{Strominger:2017zoo}
\begin{align}\label{BC-null}
    \qquad A_u\sim \mathcal{O}(1/r),\qquad A_r\sim \mathcal{O}(1/r^2),\qquad A_{z,\bar z}\sim \mathcal{O}(1),
\end{align}
which imply the following falloff behavior for the field strength
\begin{align}\label{BC-F}
    F_{ru}\sim \mathcal{O}(1/r^2),\qquad F_{rz}\sim \mathcal{O}(1/r^2),\qquad F_{uz}\sim \mathcal{O}(1),\qquad F_{z\bar z}\sim \mathcal{O}(1).
\end{align}
These falloff conditions are consistent in the sense that they include all solutions of interest, including radiation generated by localized sources. Moreover, they lead to well-defined asymptotic symmetry algebra with finite charges.

The gauge condition \eqref{Lorenz-gauge} implies that the residual gauge symmetries are solutions to $\Box \lambda=0$. 
Considering the asymptotic expansion $\lambda=\sum_{n=0}^\infty\dfrac{\lambda^{(n)}}{r^n}$,  yields 
\begin{align}
    -\dfrac{2}{r}\partial_u\lambda^{(0)}+\sum_{n=0}^\infty\dfrac{1}{r^{n+2}}\left(2n\partial_u \lambda^{(n+1)}+n(n-1)\lambda^{(n)}+\dfrac{2}{\gamma_{z\bar z}}\partial\bar\partial \lambda^{(n)}\right)=0.
\end{align}
The leading order equations are
\begin{align}\label{asymptotic-eqs}
    \partial_u\lambda^{(0)}=0,\qquad \partial\bar \partial\lambda^{(0)}=0.
\end{align}
While the leading order $\lambda^{(0)}$ is decoupled from the rest of $\lambda^{(n)}$, the subleading functions $\lambda^{(n)}\ (n>1)$ are specified in terms of $\lambda^{(1)}$, which is completely unconstrained.  
$\lambda^{(0)}$ which solves \eqref{asymptotic-eqs}  is  then
\begin{align}
    \lambda^{(0)}=f(z)+\bar f(\bar z).
\end{align}
Taking the holomorphic sector into account, the solution can be expanded as 
\begin{align}\label{LGTs-holo}
    f(z)=\alpha_0+p\ln z+\sum_{n\neq 0} \alpha_n z^n.
\end{align}
These are the LGT and are generators of non-zero soft charges
which  label points of the phase space. The subleading transformations are, however, trivial and denote degeneracy of the presymplectic form and are modded out in the physical phase space. Therefore, we concentrate only on the leading part given by equation \eqref{LGTs-holo} and its anti-holomorphic counterpart. In the following subsections, we will first compute the algebra of charges and then discuss the physical meaning of LGTs.

\subsection{Radiative phase space: symplectic and Poisson structures}

Definition of the symplectic structure at null infinity ${\cal I^+}$ is not as straightforward as the one  over spatial hypersurfaces. The reason is that null infinity is not a Cauchy hypersurface. The naive construction $\int_{\cal I^+}\omega$ fails to define a consistent symplectic form as the massive particles never reach null infinity, but instead flow through the future infinity $i^+$. To remedy this, one has to complete the symplectic form by defining it over the complete Cauchy surface $\Sigma$.  As shown in figure \ref{null-infinity-Cauchy}, this can be done in two different ways. One way is to consider instead of null infinity, a constant time surface $t=T$ and then taking the limit $T\to \infty$. The other way is to regularize the null infinity by cutting it at a sphere at large $u=U$ and attaching it to a spacelike section. That is 
\begin{align}\label{CauchySplitting}
    \Sigma=\mathcal{I}^+_{reg} \cup \Sigma',\qquad \partial\Sigma={\cal I}^+_-
\end{align}
where $\mathcal{I}^+_{reg}$ is the future null infinity truncated at a sphere $S$ at large $u=U$  and $\Sigma'$ is a spacelike section whose boundary is the same sphere $S$ at null infinity. In practice, this is splitting the phase space into radiative modes and massive modes. Accordingly, the symplectic structure is also decomposed as \cite{Strominger:2015bla,Campiglia:2015qka}
\begin{align}\label{PSSplitting}
    \bOmega^{\Sigma}=\bOmega_R+\bOmega_M
\end{align}
where
\begin{align}\label{rad-Omega}
   \bOmega_R&=\int_{\mathcal{I}^+_{reg}} \big(\df F_{uz}\wedge \df A_{\Bar{z}}+\df F_{u\Bar{z}}\wedge \df A_{z}\big) du d^2z,\\
   \bOmega_M&=-\int_{\Sigma'}\ast \df F\wedge \df A+\omega_M 
\end{align}
where $\omega_M$ is the symplectic current of the massive charged matter field. For instance, for a complex scalar, $\omega_M=\df (D\varphi)^*\wedge\df \varphi+c.c.$. Importantly, the completion of $\mathcal{I}^+$ into a Cauchy surface removes surface terms arising from the future boundary of null infinity, namely $\mathcal{I}^+_+$. An alternative way to do this without completing $\mathcal{I}^+$ into a Cauchy surface is to introduce edge modes at the boundary $\mathcal{I}^+_+$ to cancel out the boundary integrals appearing there. This was nicely formulated  in \cite{Donnelly:2016auv} and extended to different theories in \cite{Geiller:2017xad,Geiller:2017whh,Speranza:2017gxd}. 
\begin{figure}
    \centering
    \captionsetup{width=.8\linewidth}
    \includegraphics[scale=.6]{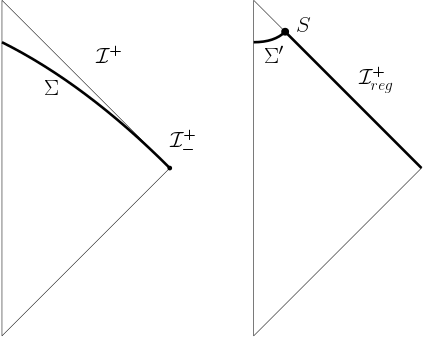}
    \caption{Different ways of turning $\mathcal{I}^+$ into a Cauchy surface. Either a) consider a constant time $t=T\to \infty$ as in the left panel, or b) cutting the null infinity at large $u$ and attaching to it a spacelike surface which extends to $r=0$ as in the right panel.}
    \label{null-infinity-Cauchy}
\end{figure}

The radiative symplectic structure $\Omega_R$ leads to the following Poisson brackets over the fields living on the null infinity, 
\be\begin{split}\label{canonical variables scri}
    \{F_{uz}(u,z,\bz),A_{\bz}(u',z',\bz')\}&=\delta(u-u')\delta^2(z-z'), \\ \{F_{u\bz}(u,z,\bz),A_{z}(u',z',\bz')\}&=\delta(u-u')\delta^2(z-z')\,,
\end{split}
\ee
which after integrating over $u$, lead to \cite{He:2014cra}
\begin{align}\label{commutator A scri}
    \{A_z(u,z,\bz),A_{\bz}(u',z',\bz')\}=\dfrac{1}{2}\Theta(u-u')\delta^2(z-z')
\end{align}
where $\Theta$ is the step function,
\begin{equation}
    \Theta(x)=\left\{
    \begin{array}{c l}
        1 &\quad x>0, \\
        -1 &\quad x<0,\\
        0 & \quad x=0.
    \end{array}\right.
\end{equation}
The value at $x=0$ is not implied by \eqref{canonical variables scri}, but given the antisymmetry of the bracket, our choice is the only way to make sense of \eqref{commutator A scri} when $u=u'$.

\subsection{Hamiltonian generators and charges}\label{HchargesNull}

The gauge transformation $A\to A+d\lambda$ generates a Hamiltonian flow over the phase space which preserves the symplectic structure. Therefore there exists a Hamiltonian function which generates this flow through the Poisson bracket. The Hamiltonian generator is given by $\df \bQ_\lambda=-\di_{\delta_\lambda}\bOmega$. It was shown by Wald \cite{Wald:1993nt, Iyer:1994ys} that for internal gauge symmetries, i.e. those under which the Lagrangian is strictly invariant, $-\di_{\delta_\lambda}\bOmega=\df J_\lambda$ where $J_\lambda$ is the Noether current associated to the gauge transformation $\delta_\lambda$. If the charged matter field $\phi$ transforms as $\phi\to e^{ie\lambda}\phi$, one can show that $J_\lambda=\lambda \mathbf{j}$ where $\bf j$ is the charge density which appears on the RHS of the Lorentz equation $d\ast F=\bf j$. Therefore we get
\begin{align}
    \df \bQ_\lambda&=-\di_{\delta_\lambda}\bOmega=-\int_\Sigma \df \ast F\wedge d\lambda+\lambda \df \mathbf{j}\cr
    &=\int_\Sigma \lambda\df (d\ast F-{\bf{j}})-\oint_{{\cal I}^+_-}\lambda \ast \df F.
\end{align}

Since the parameter $\lambda$ is field independent, the charges are manifestly integrable leading to the Hamiltonian generators  $\bQ_\lambda=\bQ_\lambda^{R}+\bQ_\lambda^{\Sigma'}$. For later use, we write the contribution of the radiative phase space $\bQ_\lambda^{R}$ explicitly
\begin{align}\label{radiative electric charge}
    \bQ_\lambda^{R}=-\int_\Sigma \ast F\wedge d\lambda=\int du d^2z (F_{uz}\partial_{\bz}\lambda+F_{u\bz}\partial_z \lambda).
\end{align}
The on-shell value of the Hamiltonian generator, the charge,  is then
\begin{align}
    Q_\lambda=-\oint_{{\cal I}^+_-}\lambda\ast F=\oint d^2z  \gamma_{z\bar z}\ \lambda  F_{ru}^{(2)}.
\end{align}
To write the last term in terms of the canonical variables we use the Lorenz gauge \eqref{Lorenz-gauge}.
Noting the boundary conditions \eqref{BC-null}, the leading order of \eqref{Lorenz-gauge} appears at $\mathcal{O}(1/r^2)$, which is
\begin{align}
    -\partial_u A_r^{(2)}-A_u^{(1)}+\frac{1}{\gamma_{z\bar z}}(\partial A_{\bar z}^{(0)}+\bar\partial A_z^{(0)})=0.
\end{align}
Next, note that $F_{ru}^{(2)}=-A_u^{(1)}-\partial_u A_r^{(2)}$ and therefore, ${\gamma_{z\bar z}}F_{ru}^{(2)}=-(\partial A_{\bar z}^{(0)}+\bar\partial A_z^{(0)})$ in the Lorenz gauge. The expression of the charge is hence,
\begin{align}
    Q^E_f=-\oint d^2z f(\partial_z A^{(0)}_{\bar z}+\partial_{\bar z} A_z^{(0)}).
\end{align}

As in the previous section, one can define the set of magnetic charges (or Hamiltonain generators) as 
\begin{align}
    \bQ^B_g=\int_{\Sigma} F\wedge dg=\oint_{{\cal I}^+_-} A\wedge dg, 
\end{align}
for $g\neq 1$ and for $g=1$ (corresponding to total magnetic charge) $\bQ^B_{g=1}=\oint_{{\cal I}^+_-} F$. Here $\Sigma$ is the Cauchy surface depicted in the left diagram in figure \ref{null-infinity-Cauchy}.
Written in $(z,\bar z)$ coordinates, and using $dz\wedge d\bz=id^2z$ we obtain
\begin{align}\label{magnetic charge-null}
    \bQ^B_g=i\oint d^2z (A_z \partial_{\bz} g-A_{\bz} \partial_z g).
\end{align}
\subsection{Algebra of charges}\label{Algebra of charges null}
Using the off-shell expressions for the charges and the Poisson brackets, we can compute the algebra of charges. Since the magnetic charge \eqref{magnetic charge-null} is written as a surface integral over $\mathcal{I}^+_-$, the relevant part of the electric charge for this computation is the contribution of the radiative phase space \eqref{radiative electric charge}. Hence for generic $f, g$ we get,
\begin{align}\label{Poisson brackets-null}
    \{\bQ^E_f,\bQ^B_g\}&=i\int dud^2z \oint d^2z' \{F_{u\bz}(x),A_{z}(x')\}\partial_z f(z)\partial_{\bz}g(z')-\{F_{uz}(x),A_{\bz}(x')\}\partial_{\bz} f(z)\partial_{z}g(z')\nonumber\\
    &=i\int dud^2z \oint d^2z' \delta(u-u')\delta^2(z-z')\Big(\partial_z f(z)\partial_{\bz}g(z')-\partial_{\bz} f(z)\partial_{z}g(z')\Big)\nonumber\\
    &=i\oint d^2z(\partial_z f\partial_{\bz}g-\partial_{\bz} f\partial_{z}g)=\oint df\wedge dg.
\end{align}
This matches with the result obtained at spatial infinity. The rest of the commutators can be also computed:
\begin{align}
    \{\bQ_{f}^E,\bQ_{g}^E\}&=\int \int du du' d^2z d^2z' \partial'_u \delta(u-u')\delta^2(z-z') \bigg[\partial f \bar{\partial'}g'+\partial_{\bz}f \partial' g' \bigg]\nonumber \\
&=\Lambda \oint d^2z(\partial_z f\partial_{\bz}g+\partial_{\bz} f\partial_{z}g),
\end{align}
where the constant $\Lambda$ is defined as 
\begin{align}
    \Lambda&=\int du\int du' \partial'_u \delta(u-u').
\end{align}
This commutator is antisymmetric, only if  $\Lambda=0$ which is indeed the case.\footnote{One can show this by replacing the delta function $\delta(x)$ by e.g. $\arctan'(\alpha x)$ and finally taking the limit $\alpha\to \infty$.} Therefore, we find $\{\bQ_{f}^E,\bQ_{g}^E\}=0$. On the other hand
\begin{align}
    \{\bQ_{f}^B,\bQ_{g}^B\}&= \oint \oint d^2z d^2z'\bigg(\{A_z(z),A_{\bz}(z')\}\partial_{\bz}f(z)\partial_{z'}g(z')+\{A_{\bz}(z),A_{z}(z')\}\partial_{z}f(z)\partial_{\bz'}g(z')\bigg).
\end{align}
Using \eqref{commutator A scri}, the same point $u=u'\to-\infty$ commutator $\{A^-_z(z),A_{\bar z}^-(z)\}=0$ and hence this commutator also vanishes. The complete algebra is hence
\begin{align}\label{AlgebraNull}
    \{\bQ^E_f,\bQ^E_g\}&=0,\qquad \{\bQ^B_f,\bQ^B_g\}=0,\qquad \{\bQ^E_f,\bQ^B_g\}=\oint df\wedge dg.
\end{align}

As a prelude to the next section, we finish this section by  exploring whether the magnetic charge generates a transformation on the radiative phase space. To this end  let us compute,
\be\begin{split}\label{QBAction}
    \{\bQ^B_g,A_z(u,z,\bar z)\}&=-i\oint d^2w \partial_w g \{A_{\bar w}(-\infty,w,\bar w),A_z(u,z,\bar z)\}=\frac{i}{2}\partial_zg,\\
    \{\bQ^B_g,A_{\bz}(u,z,\bar z)\}&=i\oint d^2w \partial_{\bar w} g \{A_{z}(-\infty,w,\bar w),A_{\bz}(u,z,\bar z)\}=-\frac{i}{2}\partial_{\bz} g.
\end{split}
\ee
One should note that the above relations have been written for $u\neq -\infty$; they are vanishing when $u=-\infty$.  
The above implies that the magnetic charge has a local action on the radiative phase space, which despite the resemblance, is not a gauge transformation of the usual form.  In section \ref{sec:4.2}, we show that this becomes a true symmetry of the theory in the duality symmetric formulation of Maxwell theory, where we add a magnetic boundary gauge transformation (denoted through $\psi_C$).

\section{Duality symmetric electromagnetism, its phase space  and  soft charges}\label{DS section}

In this section, we review our results in the more natural and electric-magnetic symmetric context of dual symmetric Maxwell theory. In this picture, the magnetic generators $\bQ^B_g$ are also generators of a $U(1)$ gauge symmetry and $Q^B_g$ are promoted to Noether charges of this symmetry. We start by  reformulation of the Maxwell theory through a dual symmetric Lagrangian \cite{Zwanziger:1968rs, Zwanziger:1970hk,bliokh2013dual,cameron2012electric}\footnote{{One could also perform a Hamiltonian analysis of the dual symmetric theory \cite{Deser:1976iy, Barnich:2007uu}. See \cite{Barnich:2007uu} for duality invariant analysis of charged black hole thermodynamics and \cite{Bunster:2018yjr} for a recent analysis of asymptotic symmetries in this context based on the Regge-Teitelboim idea \cite{Regge:1974zd}.}}. This is done by introducing another vector potential $C$ into the theory. The dual symmetric theory is governed by the Lagrangian 
\begin{eqnarray}\label{DLagrangian}
{\cal L}=-\frac12(F \wedge \ast F+G\wedge \ast G),
\end{eqnarray} 
where $F=dA$ and $G=dC$. Upon the constraints
\begin{equation}\label{Dconstraints}
\Phi=G-\ast F=0,
\end{equation}
the equations of motion of this theory becomes that of the  Maxwell theory. Meanwhile the novel symmetry structure of this theory, as we will establish, puts the electric and magnetic soft charges of previous sections on the same footing. 

We start by varying the Lagrangian,
\begin{eqnarray}
\df {\cal L}=-(d\ast F\wedge \df A +d\ast G\wedge \df C)+ d\Theta,
\end{eqnarray}
This leads to the field equations and  the presymplectic potential density $\Theta$,
\begin{align}\label{DS eom}
d \ast F=0,\quad d \ast G=0, \qquad \Theta=-(\ast F\wedge \df A+\ast G\wedge \df C).
\end{align}
Therefore, the presymplectic structure of the duality symmetric theory turns out to be 
\begin{eqnarray}\label{Symplectic-DS}
\bOmega=\int_{\Sigma} \df\Theta=\int_{\Sigma}\omega=-\int_{\Sigma}\big(\ast \df F\wedge \df A+\ast \, \df G\wedge \df C\big).
\end{eqnarray}
The covariant phase space of the duality symmetric theory is parametrized by $A_a(x;t), C_a(x;t)$. This ``extended'' phase space is twice as big as that of Maxwell theory. 

The Lagrangian as well as the constraints are invariant under the two gauge symmetry transformations of electric magnetic type
\begin{align}\label{gauge transforms}
    \delta_f:\quad A\to A+df,\qquad \tilde\delta_g:\quad C\to C+dg
\end{align}
The Hamiltonian generators associated to $\delta_{f}$ and $\tilde\delta_{g}$ are given as,
\begin{subequations}\label{DS-Ham-gen-QEB}
\begin{align}
\df \bQ^E_{f}&=-\di_{\delta_{f}}\bOmega=-\int_{\Sigma} \df \ast F\wedge df\\
\df \bQ^B_{g}&=-\di_{\tilde\delta_{g}}\bOmega=-\int_{\Sigma} \df \ast G\wedge dg\,.
\end{align}
\end{subequations}
These charges are evidently integrable over the covariant phase space of the duality symmetric theory. By construction $\bQ^E_f, \bQ_g^B$ are generators of the electric and magnetic gauge transformations, 
\be
\{\bQ^E_f, A_a(x)\}=\partial_af,\quad \{\bQ^E_f, C_a(x)\}=0,\qquad \{\bQ_g^B, C_a(x)\}=\partial_ag,\quad \{\bQ^B_g, A_a(x)\}=0,
\ee
and that 
\be
\{\bQ^E_{f},\bQ^B_{g}\}=0.
\ee
There is no surprise that above is different than the expressions of the Maxwell theory, as the brackets are defined and computed over the extended phase space. We still have to impose the constraints \eqref{Dconstraints} which reduce the theory to a duality symmetric version of Maxwell theory. The field equations and the bulk part of the symplectic structure of this theory is equivalent to Maxwell theory. At the same time, the boundary dynamics of this theory is extended by the addition of a magnetic edge mode, leading to a boundary dynamics symmetric under duality transformations. As in previous sections here we analyze this theory and its (soft) charge in the covariant phase space formulation. One may of course verify that the Hamiltonian formulation leads to the same results.

\subsection{Reduction to constrained on-shell phase space: spatial foliation}\label{sec:4.1}

To reduce duality symmetric extended phase space to the Maxwell one, we need to impose the  constraints \eqref{Dconstraints}. As discussed in the previous sections, we can study the off-shell phase space (a la Wald et al \cite{Lee:1990nz}) and then reduce to on-shell phase space by removing the bulk gauge transformations. Alternatively, we can follow Ashtekar et al method \cite{Ashtekar:1987tt,Ashtekar:1990gc}, by starting off with the solution phase space. The two methods has been shown to yield the same on-shell phase space in the end. In this subsection we present the final result and only discuss the on-shell duality symmetric phase space. In the next subsection, when we discuss the null-infinity foliation, we discuss the off-shell one too. 

To work through the construction of solution phase space and the associated symplectic structure, however, one should analyze the constraint \eqref{Dconstraints} more closely. There are two points to note here: (1) As discussed imposing the constraints amounts to imposing equations of motion and hence one should work with the on-shell phase space. To this end, as discussed in section \ref{sec:on-shell-maxwell}, in order to remove the degeneracy of the sympelctic structure, one should fix the bulk gauge transformations; leaving us with boundary (large) gauge transformations. (2) In defining the symplectic structure $\bOmega$ we need to introduce the Cauchy surface $\Sigma_t$. What then appears in $\bOmega$ is an  integration of a three-form along $\Sigma_t$. On the other hand, the constraint $\Phi_{\mu\nu}=(F+\ast G)_{\mu\nu}=0$ has components along $\Sigma_t$ and transverse to it. Explicitly, let us denote the time-like vector field normal to Cauchy surface $\Sigma_t$ by $t_\mu$ and the projector on surface $\Sigma_t$ by $P^\mu_\nu=\delta^\mu_\nu+t^\mu t_\nu$. The constraint can then be decomposed into two halves:
\be
\Phi^{\Sigma}_{\mu\nu}\equiv P^\alpha_\mu P^\beta_\nu\Phi_{\alpha\beta}=0,\qquad \Phi^t_{\nu}\equiv t^\mu\Phi_{\mu\nu}=0.
\ee
The part of the constraint relevant to the symplectic structure is $\Phi^\Sigma$. One may then show that the other half $\Phi^t=0$ is guaranteed through $\Phi^\Sigma=0$, once we impose field equations. In our analysis of the on-shell phase space, we hence only focus on the constraints along $\Sigma_t$, $\Phi^\Sigma=0$.

To impose the $\Phi^\Sigma=0$, it is convenient to choose a time coordinate $t$ and to decompose the field strengths $F$ in terms of electric and magnetic fields as in \eqref{decomp-F},
\be
F^{0a}=E^a,\qquad F^{ab}=\epsilon^{abc}B_c.
\ee
The constraint $\Phi^\Sigma$ is then written as
\be\label{constGEB}
G^{0a}=\frac12\epsilon^{abc}F_{bc}= B^a.
\ee
Note that the other half of constraint $\Phi^t=0$, takes the form 
$G^{ab}=\epsilon^{abc}F_{0c}$ and we are \emph{not} imposing that; it follows from the equations of motion.


One can solve the constraint  \eqref{constGEB} and eliminate $C$ for $A$. This will yield ``electric'' picture and is expected to bring us back to the \emph{on-shell} Maxwell theory in the bulk, while we still remain with two boundary gauge transformations, as we will see.

\paragraph{Imposing the $\Phi^\Sigma=0$ constraint on the on-shell covariant phase space.}  Let us first generalize the facilitating notation introduced in section \ref{sec:on-shell-maxwell}. We can decompose the one-form gauge fields into an exact part and a gauge invariant part:
\be\label{hatted-gauge-field}
A=\hat A+d\psi_A,\qquad C=\hat C+d\psi_C,
\ee
where $\psi_A, \psi_C$ are two scalar functions and $\hat A, \hat C$ are gauge invariant. Under gauge transformations, $\psi_A\to \psi_A+f,\ \psi_C\to \psi_C+ g$. Next, let us compute the extended symplectic structure $\bOmega$ \eqref{Symplectic-DS} over the equations of motion \eqref{DS eom} and the constraint \eqref{constGEB}. To this end, we note that in the symplectic form \eqref{Symplectic-DS}, only the spatial components of $\ast F,\ \ast G$ appear in the integral and hence one can eliminate $G^{0a}$ in terms of magnetic field, recalling \eqref{constGEB}. The symplectic structure over the constraint, denoted by $\Omega_\Phi$, then takes the form
\begin{eqnarray}\label{Symplectic-reduced}
\Omega_\Phi=\int_{\Sigma_t}\big(\ast \df E\wedge \df A+ \df B\wedge \df C\big),
\end{eqnarray}
where $B=dA=d\hat A$, as the constraints \eqref{constGEB} imply. 
As we see, $\Omega_\Phi$ is manifestly duality symmetric. 

\paragraph{Charge analysis and boundary gauge transformations, the electric picture.} 
As discussed the electric picture amounts to imposing  $\Phi^\Sigma=0$  on the on-shell $\Omega$ \eqref{Symplectic-reduced}. Explicitly, in the electric picture we substitute the ``magnetic momentum'' $G^{0a}$ in terms of $A$ ($B=dA$). The  symplectic structure on the on-shell phase space in the electric picture hence becomes
\begin{eqnarray}\label{Symplectic-reduced-E-simplified}
\Omega_\Phi=\int_{\Sigma_t}\big( \df \ast E\wedge \df \hat A+ \df dA\wedge \df \hat C\big)+ \oint_{\partial\Sigma_t} \left(\df \ast E \wedge \df \psi_A+\df A\wedge \df d\psi_C
\right),
\end{eqnarray}
From the symplectic structure ${\Omega}_\Phi$ we can compute the basic Poisson brackets, 
\begin{align}
    \{E^a(x),\hat A_b(x')\}=\{B^a(x)&,\hat C_b(x')\}=\ \delta^a_b\ \delta^3(x-x'),\qquad x,x'\in \Sigma_t,\\ 
    \{n\cdot E(x),\psi_A(x')\}= \delta^2(x-x'),\quad &\{A_i(x),\partial_j\psi_C(x')\}=\epsilon_{ij}\delta^2(x-x') \qquad x,x'\in \partial\Sigma_t,\label{E-phi-B-psi}
\end{align}
where $n$ is the vector normal to the boundary, in our case $n=dr$, and the electric and magnetic charges:
\begin{align}\label{On-shell-charges-E-DS}
 Q^E_f=\oint_{\partial\Sigma_t}   \ast E \ f,\qquad  Q^B_g= \left\{\begin{array}{cc} \oint_{\partial\Sigma_t} A\wedge dg, &\qquad {g\neq 1}\\ & \\ \oint_{\partial\Sigma_t} B= \oint_c A, & \qquad g=1,\end{array}\right.
\end{align}
The above clearly reproduces the analysis of the section \ref{sec:2} for the Maxwell theory. However, there is a very important difference: Both the electric and magnetic soft charges are now Noether charges and are associated with electric and magnetic LGTs. In other words, from \eqref{E-phi-B-psi}, one can verify that
\begin{align}
\{Q^E_f,\psi_A (x)\}=f(x),&\qquad \{Q^B_g,\psi_C(x)\}=g(x),\qquad x\in \partial\Sigma_t.
\end{align}
That is, the charges $Q^E_f, Q^B_g$ are indeed generators of boundary gauge transformations, as expected. We can now compute the algebra of charges:
\be\label{QE-QB-electric}
\{Q^E_{f},Q^B_{g}\}=\Omega_{\Phi}(\delta_f,\delta_g)=\delta_f Q^B_{g}=\oint df\wedge dg=-\oint_c gdf.
\ee

\paragraph{Charge analysis in the magnetic picture.} Alternatively, one could have taken the constraint as $ G=\ast F$ and hence used $\Phi^t=0$ half of the constraints. This amounts to eliminating the ``electric'' degrees of freedom for magnetic ones. Explicitly, in the magnetic picture we replace the ``electric momentum'' $E$ in terms of $C$ ($\ast E=-dC$). 

\paragraph{Notation.} To distinguish the on-shell quantities in magnetic picture from the electric ones, we use the following notation:
The quantity $X$ in the electric picture will be denoted by $\tilde X$ in the magnetic picture. 

The  symplectic structure on the on-shell phase space in the magnetic picture hence becomes
\begin{eqnarray}\label{Symplectic-reduced-B-simplified}
\tilde\Omega_\Phi=-\int_{\Sigma_t}\big( \df \ast G\wedge \df \hat C +\df dC\wedge \df \hat A\big)- \oint_{\partial\Sigma_t} \left(\df \ast G \wedge \df \psi_C+\df C\wedge \df d\psi_A
\right).
\end{eqnarray}
Yielding the following charges in the magnetic picture:
\begin{align}\label{On-shell-charges-B-DS}
 \tilde Q^E_f= \left\{\begin{array}{cc} -\oint_{\partial\Sigma_t} C\wedge df, &\qquad {f\neq 1}\\ & \\ \oint_{\partial\Sigma_t} *E= -\oint_c C, & \qquad f=1,\end{array}\right.,\qquad  \tilde Q^B_g=-\oint_{\partial\Sigma_t} \ast G\ g.   
\end{align}
The algebra of charges in the magnetic picture turns out to be 
\be\label{QE-QB-magnetic}
\{\tilde Q^E_{f},\tilde Q^B_{g}\}=\tilde\Omega_\Phi(\delta_f,\delta_g)=-\tilde{\delta}_g \tilde Q^E_{f}=-\oint_{\partial\Sigma_t} df\wedge dg=-\oint_c fdg.
\ee
The minus sign compared to \eqref{QE-QB-electric} is stemming from the fact that the role of $Q^E$ and $Q^B$ are exchanged in the magnetic picture. This may be seen comparing the expression of the charges in the two pictures, \eqref{On-shell-charges-E-DS} and \eqref{On-shell-charges-B-DS}.  These two pictures provide two different phase space coordinates and basis for expanding the soft charges. We will show later in this section that this exchange of pictures can be also derived from \eqref{QE-QB-electric} after performing a $\frac{\pi}{2}$ rotation under the duality symmetry transformation.

We comment that the expression of the value of two charges 
in different pictures are not exactly the same:
\begin{align}\label{bQB-vs-QB}
    \tilde Q^B_g=-\oint_{\partial\Sigma_t} \ast G\ g\stackrel{\Phi}{=}\oint_{\partial\Sigma_t} B\ g=\oint A\wedge dg +\oint d(A g) =Q^B_g +\oint d(A g)\,, \quad g\neq 1.
\end{align}
For $g=1$, $\tilde Q^B_{g=1}=Q^B_{g=1}=\oint dA$. That is the total magnetic charge (and similarly for the total electric charge) is the same in electric and magnetic pictures, as physically expected. 
The last term in \eqref{bQB-vs-QB} is vanishing if $gA$ has no residue at the possible singular point $z=0$. This is because in the presence of singularities, this integral can be written as a contour integral around the singularities. 

Interpreting the singularities as source for magnetic/electric surface multipoles, then this term is in fact a measure of these sources.  
One can use the above to show that all the charges in the two pictures commute
\begin{align}\label{E-B-mixed-algebra}
    \{Q^X_f,\tilde Q^Y_g\}&=0\,,
\end{align}
where $X,Y$ can be either $E$ or $B$. This will be of importance when we discuss the duality transformations in section \ref{sec: Duality charge}.
\subsection{Symplectic structure and soft charges at null infinity}\label{sec:4.2}

In this section, we extend the analysis of the section \ref{HchargesNull} to the duality symmetric version of the Maxwell theory given by (\ref{DLagrangian}). This analysis will shed further light on the role of the extra (magnetic) boundary degree of freedom in the duality symmetric Maxwell theory. Also, we will see how the constraints reduce to local relations at null infinity unlike the case of spatial foliation discussed earlier.

As in the previous subsection, we use equation \eqref{hatted-gauge-field} to decompose the gauge fields into an exact part and a gauge invariant parts. However, due to the boundary condition $A_u,C_u\sim \mathcal{O}(1/r)$, we find that $\psi_A,\psi_C$ are $u$ independent
\begin{align}\label{decomposition-null}
    A(u,z,\bz)=\hat A(u,z,\bz)+d\psi_A(z,\bz),\qquad C(u,z,\bz)=\hat C(u,z,\bz)+d\psi_C(z,\bz)
\end{align}
Substituting in the symplectic structure and using the equations of motion $d\ast F=0, d\ast G=0$ we arrive at
\begin{align}
    \Omega=-\int_{\Sigma}\df \ast F\wedge \df \hat A +\df \ast G\wedge \df \hat C-\oint_{\partial\Sigma} \df \ast F\wedge \df \psi_A+\df \ast G\wedge \df \psi_C
\end{align}
\paragraph{The electric picture.} By imposing the constraints $\Phi=\ast G+F$ and eliminating $C$ for $A$, we adapt the electric picture. Note that as in previous section, only half of the constraints naturally appear in the above integral. The constrained symplectic form is 
\begin{align}
    \Omega_\Phi=-\int_{\Sigma}\df \ast F\wedge \df \hat A -\df F\wedge \df \hat C-\oint_{\partial\Sigma} \df \ast F\wedge \df \psi_A-\df  F\wedge \df \psi_C
\end{align}
In components, the constraints used are $\Phi^\Sigma=\{\Phi_{uz},\Phi_{u\bz},\Phi_{z\bz}\}$. These constraints can be also used to trade $C$ for $A$. In particular, recalling equation \eqref{starF} the first two imply
\begin{align}
  \partial_u(C_z-iA_z)=0,\qquad\partial_u(C_{\bz}+iA_{\bz})=0
\end{align}
Given the decomposition \eqref{decomposition-null}, these lead to constraints between the hatted parts\footnote{That is, $C_z(u,z,\bz)=i A_z(u,z,\bz)+ iD_z(z,\bz),\  C_{\bz}(u,z,\bz)=-i A_{\bz}(u,z,\bz)-i D_{\bz}(z,\bz)$, where $D$ is a $u$-independent 1-form. Using the Hodge decomposition theorem on the sphere, $D$ can be decomposed as $D=d\alpha+\ast d\beta$ where $\alpha, \beta$ are functions on the sphere. In components $D_z=\partial_z\alpha+i\partial_z\beta$, $D_{\bz}=\partial_{\bz}\alpha-i\partial_{\bz}\beta$. These two functions however, can be absorbed into the exact parts of gauge fields i.e. $\psi_A,\psi_C$.}
\begin{align}\label{C vs A}
    \hat C_z(u,z,\bz)=i \hat A_z(u,z,\bz), \qquad \hat C_{\bz}(u,z,\bz)=-i \hat A_{\bz}(u,z,\bz).
\end{align}
Using these in the symplectic form, we arrive at
\begin{eqnarray}\label{preSymplectic-Reduced-Null}
\Omega_{\Phi}=2\int_{\mathcal{I}^+}du d^2z\Big(\df F_{uz}\wedge \df \hat A_{\bz} + \df F_{u\bz}\wedge \df \hat A_{\bz} \Big)
-\oint_{S^2}\Big(\df \ast F\wedge  \df \psi_A-\df A\wedge\df d\psi_C\Big) ,
\end{eqnarray}
where in the last term we have used an integration by parts. We note that while the magnetic gauge field $C$ and its momentum conjugate have been substituted for the electric gauge field $A$ and while the bulk part of \eqref{preSymplectic-Reduced-Null} is gauge invariant, there are two $U(1)$ boundary gauge transformations, manifested in the $\psi_A,\psi_C$ terms in the above. Let us write the boundary part in components 
\begin{align}
    \Omega_b&=\oint_{S^2}d^2z \gamma_{z\bz}\df F_{ru}^{(2)}\wedge \df \psi_A+i\oint_{S^2}d^2z(\df A_z\wedge\df\partial_{\bz}\psi_C-\df A_{\bz}\wedge\df\partial_{z}\psi_C)
\end{align}
The Hamiltonian generators for the boundary electric and magnetic gauge transformations can then be computed using \eqref{preSymplectic-Reduced-Null}:
\begin{align}
   Q^E_f=-\oint_{{\cal I}^+_-} f\ast F=\oint_{{\cal I}^+_-}d^2z f \gamma_{z\bz} F_{ru}^{(2)}\,, 
   \qquad Q^B_g=\oint_{{\cal I}^+_-}A \wedge dg=i\oint_{{\cal I}^+_-}d^2z(A_z\partial_{\bz}g-A_{\bz}\partial_{z}g).
\end{align}
We note that, as before, $Q^B_{g=1}$ different than above, and is given by $Q^B_{g=1}=\oint_{{\cal I}^+_-} F$.

To further analyze the charges and their algebra we need the basic Poisson brackets which may be read off from \eqref{preSymplectic-Reduced-Null}:
\begin{align}
     \{ F_{uz}(u,z,\bz),\hat{A}_{\bz}(u',z',\bz ')\}&=\{F_{u\bz}(u,z,\bz),\hat{A}_{z}(u',z',\bz ')\}=2\delta(u-u')\delta^2(z-z')\,,\\
     \{ F_{ru}^{(2)}(z,\bz),\psi_A(z,\bz)\}&=\delta^2(z-z'),\qquad \{A_z(z,\bz),\partial_{\bz}\psi_C(z,\bz)\}=-i\delta^2(z-z').
\end{align}
Using the Poisson brackets, one can check that the above expressions correctly generate the boundary gauge transformations 
\begin{align}
\{Q^E_f,A_i (x)\}=\partial_i f(x),&\qquad \{Q^B_g,C_i(x)\}=\partial_i g(x),\qquad x\in {\cal I}^+_- , \;i\in(z,\bz).
\end{align}
Moreover, the Poisson brackets between electric and magnetic charges yields the same as before
\begin{align}
   \nonumber\{Q^E_f,Q^B_g\}&=\oint\oint d^2z d^2w f(w,\Bar{w})\big( \{F_{ru}(w,\Bar{w}),\partial_z\psi_A(z,\bz)\} \partial_{\bar{z}}g-\{F_{ru}(w,\Bar{w}),\partial_{\bz}\psi_A(z,\bz)\} \partial_{z} g \big)\\
   &=i\oint_{{\cal I}^+_-} (\partial_{z} f \partial_{\bz}g-\partial_{\bz} f \partial_{z}g)d^2z=\oint_{{\cal I}^+_-} df \wedge dg.
\end{align}

\subsection{Duality generating charge}\label{sec: Duality charge}
One of the byproducts of taking the dual-symmetric version of the Maxwell theory is emergence of a new continuous global  symmetry $U(1)_\theta$ which rotates electric and magnetic fields into each other. We note that the Lagrangian (\ref{DLagrangian}) and the constraint (\ref{Dconstraints}) are invariant under the $U(1)_\theta$ transformation
\begin{eqnarray}\label{Duality-rotation-F,G}
F \rightarrow F \cos \theta + G \sin \theta\,,\,\,\,\,\, \,\,\, G \rightarrow G \cos \theta - F \sin \theta,
\end{eqnarray}
which in terms of $A$ and $C$,
\begin{eqnarray}\label{Duality-rotation-A,C}
A \rightarrow A \cos \theta + C \sin \theta\,,\,\,\,\,\, \,\,\, C \rightarrow C \cos \theta - A \sin \theta,
\end{eqnarray}
up to a gauge transformation. 

It can be shown that this vector field which is tangent to 
the constraint $\Phi$, is actually (pre)symplectomorphism of $\Omega$. The duality symmetry generator $\delta_\theta$ then acts on fields as
\begin{align}\label{duality-infinitesimal}
\delta_{\theta}(A,C)=(C,-A),\qquad \delta_{\theta}(F,G)=(G,-F).
\end{align}

\paragraph{Off-shell duality symmetry and its charge.}

Denoting the  Lie derivative with respect to the vector $\delta_{\theta}$ in the space of fields by  $\mathbbmss{L}_{\delta_\theta}$, it is readily seen that $\mathbbmss{L}_{\delta_\theta}\bOmega=0$, with $\bOmega$ given in \eqref{Symplectic-DS}.
Generator of the duality-symmetry symplectomorphism is the duality charge $\bQ_{\theta}$ (which, for the reasons becoming clear in the next subsection, is also called optical helicity \cite{cameron2012electric})  is computed as
\begin{eqnarray}
\df \bQ_{\theta}=-\di_{\delta_{\theta}}\bOmega=\int_{\Sigma_t}[\df(\ast G\wedge A)-\df(\ast F\wedge  C)]\,.
\end{eqnarray}
As is manifestly seen the above charge is integrable and hence we find the generator of the duality transformation as 
\begin{align}\label{bQ-theta}
\bQ_{\theta}=\int_{\Sigma_t}d^3x\sqrt{|h|} (F^{0a} C_a-G^{0a} A_a)\,.
\end{align}
The above is nothing but the standard Noether charge associated with $U(1)_\theta$. 

The algebra between $\bQ_\theta$ and electric and magnetic soft charges can be computed using the Poisson brackets deduced from (\ref{Symplectic-DS}),
\begin{align}\label{Qtheta-Q-algebra}
\{\bQ_{\theta},\bQ^E_f\}=\bQ^B_f\,, \qquad \qquad \{\bQ_{\theta},\bQ^B_f\}=-\bQ^E_f\,.
\end{align}
This algebra may be written in the \eqref{complex-basis-f} basis:
\be\label{Qtheta-Qn-algebra}
\begin{split}
&\{\bQ_\theta, \bQ_n^{E} \}=\bQ_n^{B} \,, \qquad \{\bQ_\theta,\bQ_n^{B}\}=-\bQ_n^{E}\,, \\
&\{\bQ_\theta,\mathbf{P}^{E}\}=\mathbf{P}^{B}\,,\qquad   \;\{\bQ_\theta,\mathbf{P}^{B}\}=-\mathbf{P}^{E}. 
\end{split}
\ee
This algebra was also discussed in \cite{Bhattacharyya:2017obx} and has infinite $iso(2)$ sub-algebras for any given $n$. Similar, but not exactly identical, $iso(2)$ algebras was also discussed in \cite{Hamada:2017bgi}. We shall make further comments on the latter below in this section. 

\paragraph{Optical helicity operator $Q_\theta$.} As the next step, we impose the constraint \eqref{Dconstraints} which also amounts to going on-shell. Alternatively, one can impose equations of motion and $\Phi^\Sigma$ \eqref{constGEB}. $\Omega_\Phi$ \eqref{Symplectic-reduced} is formally invariant under
duality symmetry transformation. However, one should note that this is at formal level; depending on whether we are in electric or magnetic pictures, respectively either $B=dA$ or $\ast E=-dC$, we lose this ``manifest'' duality.
The above may be put in a different wording: Sympletic structure of the duality symmetric on-shell phase space could be represented in different basis, \eqref{Symplectic-reduced-E-simplified} or \eqref{Symplectic-reduced-B-simplified}. While the phase space itself and hence the set of soft charges are invariant under $\theta$-rotations, the phase space coordinate used is not. Explicitly, $Q_\theta$ is the generator of this coordinate transformation on the phase space. In the particular case of electric of magnetic basis, it is evidently seen that \eqref{Symplectic-reduced-E-simplified} and \eqref{Symplectic-reduced-B-simplified} rotate into each other. Nonetheless, as we will show below, the algebra of charges does remain duality invariant irrespective of the basis used, as expected.

The duality charge $Q_\theta$ is then given by the same expression as in \eqref{bQ-theta}, explicitly
\begin{align}\label{Q-theta}
Q_{\theta}=\int_{\Sigma_t}d^3x\sqrt{|h|} ( E^a C_a-B^a A_a)\,,
\end{align}
where in the electric picture $G^{0a}=B^a\equiv (\nabla\times A)^a$, while in the magnetic picture $F^{0a}\equiv E^a=-(\nabla\times C)^a$. Therefore, under the duality transformation \eqref{Duality-rotation-F,G}, $\delta_\theta(\vec{E}, \vec{B})=(\vec{B}, -\vec{E}).$
{It is well known that $Q_\theta$ measures the total helicity, namely the difference of the number of right handed and left handed photons \cite{Deser:1976iy}. This can be understood by expanding the field in plane waves, and observing that the duality transformation is  indeed a rotation in the transverse plane for each wave. Accordingly the corresponding charge for each mode coincides with its helicity.} This will become more explicit in the end of section \ref{sec:4.4}.

One can then compute the algebra of charges in electric picture, using \eqref{Symplectic-reduced-E-simplified}, or in magnetic picture, using \eqref{Symplectic-reduced-B-simplified}. 
As discussed commutation with $Q_\theta$ besides changing electric charges to magnetic (and vice versa), also changes the electric picture to magnetic one. Explicitly, in terms of the notation introduced in section \ref{sec:4.1}, 
\be\label{Qtheta-QEB-algebra}
\begin{split}
&\{Q_\theta, Q_n^{E} \}=\tilde Q_n^{B} \,, \qquad \;\;\, \{Q_\theta,{P}^{E}\}=\tilde{P}^{B}\,, \\
&\{Q_\theta,Q_n^{B}\}=-\tilde Q_n^{E}\,,\qquad   \{Q_\theta,{P}^{B}\}=-\tilde{P}^{E},
\end{split}
\ee
where we used the $f=z^n, \ \ln z$ basis. We have  similar Poisson brackets between $Q_\theta$ and charges in the magnetic picture $\tilde Q_f$. 

Using the charge algebra \eqref{QE-QB-electric}, \eqref{QE-QB-magnetic} and \eqref{E-B-mixed-algebra} one can represent $Q_\theta$  in terms of
$Q$'s and $\tilde Q$'s as
\be\label{Q-theta-electric}
Q_\theta=\frac{1}{2\pi i}(Q^E_0\tilde P^E+\tilde Q^E_0 P^E -Q^B_0\tilde P^B- \tilde Q^B_0 P^B)+ \sum_{n\neq 0}\frac{1}{2\pi i n} (Q_{-n}^B\tilde Q^B_n-Q_{-n}^E\tilde Q^E_n).
\ee
That is, $Q_\theta$ in \eqref{Q-theta-electric} reproduces $\{Q_\theta, Q^E_f\},\ \{Q_\theta, Q^B_g\}$ commutators. This expression and \eqref{QE-QB-magnetic} may be used to verify that $\{Q_\theta, \tilde Q_n^{E} \}= Q_n^{B}, \{Q_\theta, \tilde Q_n^{B} \}= -Q_n^{E}$. This is consistent with the picture that the role of $Q^E$ and $Q^B$ are exchanged in the electric and magnetic pictures. One should note that the expression \eqref{Q-theta-electric} only captures the part of $Q_\theta$ which satisfies \eqref{Qtheta-QEB-algebra}; $Q_\theta$ may in general have a part which commutes which the charges which does not appear in \eqref{Q-theta-electric}.

We close this subsection by remarking that the set of three charges $(Q_\theta, Q^E_f, \tilde Q^B_f)$, and likewise $(Q_\theta, \tilde Q^E_f,  Q^B_f)$, for any given $f$, form an $iso(2)$ algebra. That is, we have infinitly many $iso(2)$ algebras. As we see the compact part of these $iso(2)$'s is the duality charge generator $Q_\theta$ and the non-compact parts are the electric and magnetic soft charges. While resembling the algebra discussed in \cite{Hamada:2017uot, Hamada:2017bgi}, this $iso(2)$ is not exactly the same.  We shall discuss this point further in the end of next subsection \ref{sec:4.4}.

\subsection{Poincare generators, the electric-magnetic soft charges and their algebra}\label{sec:4.4}
The theory we are considering, the Maxwell theory or its dual symmetric version on four dimensional flat space, besides the gauge symmetries and the $U(1)_\theta$ global symmetry discussed above, has other global symmetries associated with the background Minkowski spacetime. These are the 10 generators of the Poincar\'e algebra, which are generated by Killing vectors $\xi$. Following \cite{iyer1994some} the variation of Poincare generators $\bQ_\xi$ is as follows,
\begin{align}
\df \bQ_\xi=-\di_{\delta_\xi} \bOmega=\int_{\Sigma} \df J_\xi-\int_{i^0} \xi \cdot \Theta\,.
\end{align}
Adopting the boundary conditions \eqref{BC1} and \eqref{BC3} at $i^0$, the boundary term $\xi \cdot \Theta$ dose not contribute\footnote{This may be explicitly verified by writing the ten killing vector fields in the $(t,r,z,\bar{z})$ coordinates, for example $(\frac{\partial}{\partial \mathrm{x}_1})_{\mu}=\frac{1}{1+z\bar{z}}(0,z+\bar{z},r,r)$, computing the integral of $\xi \cdot \Theta$ on a $t,r=const.$ surface, and taking  the $r \rightarrow \infty$ limit using  falloff conditions \eqref{BC1} and \eqref{BC3}. Notice that for those vector fields which are tangent to the surface $t,r=const.$ like rotations, the pull back of $\xi \cdot \Theta$ to the surface is vanishing and so its integral is manifestly zero.}    and  we have,
\begin{align}
\bQ_\xi=\int_{\Sigma}J_\xi=-\int_{\Sigma} (\ast F \wedge {\cal L}_\xi A+\ast G \wedge {\cal L}_\xi C) - \frac{1}{2}\xi \cdot (F \wedge \ast F+G \wedge \ast G)\,,
\end{align} 
where ${\cal L}_\xi$ denotes the Lie derivative along $\xi$ and $J_\xi=\Theta(\delta_\xi)-\xi \cdot \mathcal{L}$ is the Noether current associated with  $\xi$'s. This Noether current can also be constructed explicitly using the standard Noether procedure, leading to  $dJ_\xi=T^{\alpha \beta} {\cal L}_\xi g_{\alpha \beta}=0$, where $T^{\alpha\beta}$ is the symmetrized energy-momentum tensor of the theory and ${\cal L}_\xi g_{\alpha\beta}$ denotes the Lie derivative of the background Minkowski metric along $\xi$. This latter vanishes as $\xi$ are isometries of the background.  This leads to conservation of Poincare charges which is achieved by the boundary condition at $i^0$. 

Given expression of the Hamiltonian generators one can compute the charge algebra:
\be\begin{split}\label{PEMalgabra}
\{\bQ_\xi,\bQ_f^{E}\}=-\bQ^{E}_{{{\cal L}_\xi} f}\,, \qquad \{\bQ_\xi,\bQ_g^{B}\}&=-\bQ^{B}_{{{\cal L}_\xi} g}\,, \\
\{\bQ_{\xi_{1}},\bQ_{\xi_{2}}\}=\bQ_{[\xi_{1},\xi_{2}]}\,, \qquad \{\bQ_\xi,\bQ_\theta\}&=0\,,
\end{split}\ee
where $[\xi_{1},\xi_{2}]={\cal L}_{\xi_1}\xi_2=-{\cal L}_{\xi_2}\xi_1$ gives the Poincare algebra. 
We note that in the analysis above we did not crucially use $\xi$'s to be Poincar\'e generators, for most of our analysis $\xi$ could be any diffeomorphism which keep Maxwell action invariant and respect mild boundary falloff behavior needed for our charge analysis. A similar analysis in the Hamiltonian formulation has been recently carried out in \cite{Henneaux:2018gfi,Henneaux:2018hdj}. The algebra \eqref{PEMalgabra} is only a manifestation of the fact that the soft charges are scalars, depending on a scalar function $f$ and that the duality charge $\bQ_\theta$ is a scalar over the spacetime. In particular, for the Maxwell theory one may consider $\xi$ to be the conformal Killing vectors, generating the conformal group $SO(4,2)$.  

One should note that these Hamiltonian generators will become conserved charges once computed on-shell and that these on-shell charges act as generators of associated transformations on the on-shell phase space consisting of physical transverse as well as the boundary (soft) photons. The charge $\bQ_\xi$ are computed for the duality symmetric theory \eqref{DLagrangian}. For the on-shell Maxwell theory, besides the equations of motion we need to impose the constraint $\Phi^\Sigma=0$. Using the usual  decomposition into $E$ and $B$ and choosing the Cauchy surfaces $\Sigma_t$, we have,
\begin{align}\label{Q-xi-general}
Q_\xi=
\int_{\Sigma_t} (\ast E \wedge {\cal L}_\xi A+B \wedge {\cal L}_\xi C)=\int_{\Sigma_t} d^3x\sqrt{|h|} (E^a ({\cal L}_\xi A)_a+B^a ({\cal L}_\xi C)_a)\,,
\end{align}
 where as previous sections, depending on choosing electric or magnetic picture, $B=dA$ or $E=-dC$, respectively. 

One may then check the algebra of $Q_\xi$ with electric or magnetic charges in the electric \eqref{QE-QB-electric} or magnetic \eqref{QE-QB-magnetic} frame and also with duality symmetry charge $Q_\theta$ \eqref{Q-theta}. In the electric frame a straightforward computation using electric  symplectic structures \eqref{Symplectic-reduced-E-simplified} yields 
\be\begin{split}\label{PEMalgabra-onshell}
\{Q_\xi,Q_f^{E}\}=-Q^{E}_{{{\cal L}_\xi} f}\,, \qquad \{Q_\xi,Q_g^{B}\}&=-Q^{B}_{{{\cal L}_\xi} g}\,, \\
\{Q_{\xi_{1}},Q_{\xi_{2}}\}=Q_{[\xi_{1},\xi_{2}]}\,, \qquad \{Q_\xi,Q_\theta\}&=0\,.
\end{split}\ee
In the above we have used the fact that ${\cal L}_\xi (n_a) f$ where $n_a=dr$ is the normal vector to the celestial sphere is of order $(1/r)$ and hence does not contribute. Similarly one may compute the algebra in magnetic frame using \eqref{Symplectic-reduced-B-simplified}.

\paragraph{Construction of angular momentum ${\cal J}$ in terms of soft charges.} One may try to give a representation of the Poincar\'e charges $Q_\xi$ appearing in \eqref{PEMalgabra-onshell} in terms of the soft charges $Q^E, Q^B$. While the algebra \eqref{PEMalgabra-onshell} contains all of Poincar\`e generators on the same footing, the electric and magnetic soft charges are functions of $f$ which is defined on the celestial sphere. To make the analysis simpler we hence only focus on two of the Poincar\'e generators which respect the asymptotic decomposition of the spacetime; among the ten $Q_\xi$ we only consider the one associated with energy, $\xi=\partial_t$ and the one associated with ``angular momentum'' $\xi=-i(z\partial_z -\bz\partial_{\bz})$; the generators of these will be respectively denoted by ${\mathcal{H}}$ and  ${\mathcal{J}}$.\footnote{ It is worth noting that \eqref{Q-xi-general} reduces to the familiar expressions of the energy  for $\xi=\partial_t$ and to spin angular-momentum of electromagnetic field for rotations. For $\xi=\partial_t,\ {\cal L}_\xi A_a=\partial_t A_a=E_a+\nabla_a\Phi_E$ and ${\cal L}_\xi C_a=\partial_t C_a=B_a+\nabla_a\Phi_B$ where $\Phi_E, \Phi_B$ are electric and magnetic scalar potentials. Plugging these into \eqref{Q-xi-general}, we obtain the ``total'' Hamiltonian $\mathcal{H}=\int_{\Sigma_t} [(E^2+B^2)+\vec{\nabla}\cdot\vec{E} \Phi_E+\vec{\nabla}\cdot\vec{B}\Phi_B]$. Similarly, for the three rotation generators $\xi^{(a)}=\epsilon^{abc}x_b\partial_c$, if we take the internal part of ${\cal L}_\xi A_a$ i.e. $A^b\partial_a \xi^b$ and similarly for the magnetic counterpart, we obtain $\vec{\mathcal{S}}=\int_{\Sigma_t} (\vec{E}\times\vec{A}+\vec{B}\times\vec{C})$ which gives the ``spin'' (non-orbital) part of the angular momentum. It is manifestly seen that these expressions for $\mathcal{H}, \vec{\mathcal{S}}$ are invariant under electric-magnetic duality transformation \eqref{Duality-rotation-F,G},\eqref{Duality-rotation-A,C}.\label{footnote-spin}}

For these on-shell boundary generators and in the basis \eqref{complex-basis-f} for $f$, the algebra takes the form:
\be\begin{split}\label{H-Q-algabra}
\{{\mathcal H},Q_n^E\}=0\,&, \qquad \{{\mathcal{H}},Q_n^{B}\}=0\,, \\
\{{\mathcal{H}},{P}^E\}=0\,&, \qquad \{{\mathcal{H}},{P}^B\}=0\,.
\end{split}\ee
The above confirms that $Q_n^E, {P}^E$ and their magnetic counterparts $ Q_n^B, {P}^B$ are indeed soft charges, as they commute with the Hamiltonian. Or, alternatively the above confirms conservation of these charges. The commutators involving $\mathcal{J}$ are
\be\begin{split}\label{JQalgebra}
\{{\mathcal{J}}, Q_n^{E} \}=inQ_n^{E} \,&, \qquad \{\mathcal{J},{P}^{E}\}=iQ_0^{E}\,,\\
\{{\mathcal{J}}, Q_n^{B} \}=in Q_n^{B} \,&, \qquad \{\mathcal{J},{P}^{B}\}=iQ_0^{B}\,,
\end{split}\ee
As \eqref{JQalgebra} indicates the electric and magnetic charges $Q_0^E, Q_0^B$ commute with ${\mathcal{J}}$. 
As in the $Q_\theta$ case, one may try to represent $\mathcal{J}$ in terms of $Q^E, Q^B$. One may easily check that\footnote{We are reading the expression ${\cal J}$ \eqref{S-QE-QB} from the commutation relations \eqref{JQalgebra}. In principle the angular momentum ${\cal J}$ may have a part which commutes with the soft charges. Our expression \eqref{S-QE-QB} does not capture this latter. Existence of this part, however, does not alter our discussions.} 
\be\label{S-QE-QB}
\mathcal{J}=\frac{1}{2\pi }\sum_{n} Q_{-n}^E Q_n^B,
\ee
satisfies the algebra \eqref{JQalgebra} and its counterpart in magnetic picture.  To verify commutation of spin operator and $Q_\theta$, one should introduce a ``total'' angular momentum operator which acts both on $Q$'s and $\tilde Q$'s, i.e. ${\cal J}_{total}={\cal J}+ \tilde{{\cal J}}$.
Recalling \eqref{QE-QB-magnetic} one can  show that,
$$\tilde{\cal J}=-\frac{1}{2\pi }\sum_{n}\tilde Q_{-n}^E \tilde Q_n^B,$$ 
and hence $\{{\cal J}_{total}, Q_\theta\}=0$. 

As a side comment we note that ${\cal J}$ is a component of spin operator and recalling Bohr quantization, it is  semiclassically  quantized in units of $\hbar$. In particular, the zero mode of ${\cal J}$,
${\cal J}_0=\frac{1}{2\pi}Q_0^EQ_0^B$ is quantized:
\be
Q_0^EQ_0^B=2\pi n\hbar, \quad n\in \mathbb{Z}.
\ee
The above is remarkably just the usual Dirac quantization of the electric or magnetic charge.

\paragraph{Optical helicity versus spin.} {The charge $Q_\theta$ and the spin are not totally independent. To see this we note that  the integrand of \eqref{bQ-theta} after imposing the constraint $\ast G=-F$ reveals the so called \textit{helicity-spin} conserved current $J_\theta^\mu=(h,\boldsymbol{s})$ \cite{cameron2012optical,cameron2012electric,bliokh2013dual}
\begin{align}
    \partial_\mu J_\theta^\mu=0 ,\qquad h=E\cdot C-B\cdot A,\qquad \boldsymbol{s}=E\times A+B\times C 
\end{align}
While the time component is the density of the optical helicity $Q_\theta$, the spatial component is nothing but the density of the spin ${\boldsymbol{s}}$, i.e. the internal part of the angular momentum (\emph{cf.} footnote \ref{footnote-spin}). Integrating the conservation law over a spacetime region implies 
\begin{align}\label{Smu}
    \dfrac{d}{dt}Q_\theta=-\oint_{B} \boldsymbol{s}\,\cdot \,\vec n\, da
\end{align}
Accordingly, each photon escaping the boundary $B$ of the region, reduces $Q_\theta$ by the value of $\boldsymbol{s}\cdot n$, i.e. its helicity. This is notable as ${\boldsymbol{s}}$ is only the internal part of the angular momentum and its sum with the orbital part reveals another conserved quantity.  Moreover, as argued in our setting $Q_\theta$ is a conserved charge which dovetails with \eqref{Smu} recalling that with our falloff behavior $\oint \vec{n}\cdot \boldsymbol{s}=0$.
}

\paragraph{More on the $iso(2)$ algebras.} In \cite{Hamada:2017bgi} two sets of $iso(2)$ algebras were discussed; one is {the quantum version of our classical results at} 
 the end of section \ref{sec: Duality charge}\footnote{Note that this $iso(2)$ in \cite{Hamada:2017bgi} is written in terms of creation-annihilation operators of photons reaching ${\cal I}^+$.}  and the other is associated with the little group of the Poincare group for photons. These two $iso(2)$ algebras were then discussed to be identical. Here we argue that this latter cannot be true. 
Consider a single photon state of frequency $\omega$  moving in direction $\vec{k}$ in the radiation gauge. It is straightforward to show that for this state $\vec{k}.(\vec{E}\times \vec{A})=\omega\vec{B}\cdot \vec{A}$. That is, density of $Q_\theta$ \eqref{Q-theta} and the helicity density of the photon ($\vec{k}\cdot \vec{s}/\omega$) are equal to each other. Next, we note that this expression is zero for a linearly polarized photon and its value for clockwise and counterclockwise circularly polarized photons differs by a sign. Therefore, by superposition, $Q_\theta$ for a system of photons  measures the difference between number of two circular polarizations. We also learn that for a generic system of photons the angular momentum ${\cal J}$ is not equal to  $Q_\theta$. This discussion implies that, while the compact generator in both of the  $iso(2)$ algebras discussed in \cite{Hamada:2017bgi} is the optical helicity (i.e. expression \eqref{Q-theta} reduces to (2.18) in \cite{Hamada:2017bgi}), the non-compact ones, i.e. $Q^E_f, Q^B_f$, are not related to the $iso(2)$ generators associated with the little group of Poincare group for massless photons. One simple reason to see this is that the electric and magnetic soft charges are linear in the gauge field $A$ and its time or space derivatives, whereas the Poincare generators are quadratic in fields. It is known that non-compact $iso(2)$ generators of the little group act on  photon fields as gauge transformations \cite{Weinberg:1964ew}, nonetheless, the parameter of this gauge transformation is field dependent (it is linear in $A$) and is not among the set of functions $f, g$ we considered here.

\section{Discussion and outlook}\label{discussion:sec}
In this paper we analyzed the soft charges of Maxwell theory, which has been extensively studied in the literature, further focusing on the magnetic soft charges. In particular, we analyzed the charges as functions over the phase space of the theory and computed their Poisson brackets, allowing for gauge transformations which are singular at the celestial sphere. We found that while electric soft charge (and magnetic soft charges) commute and form an Abelian algebra the magnetic and electric charges do not commute. Of course the non-Abelian algebra appears if we allow the gauge transformations which have localized mild singularity on the celestial sphere at infinity. These gauge transformations are typically the ones which are used in the soft charge analysis \cite{Strominger:2017zoo}, and/or in the similar analysis in gravity, yielding BMS algebras \cite{Barnich:2010eb, Barnich:2011mi, Barnich:2017ubf}. 
Here we would like to discuss some of the physical implications of our algebras and some possible future directions.

\paragraph{Physical interpretation of large gauge transformations.}
Besides $Q_0^E, Q^B_0$ charges which correspond to electric or magnetic monopole charges and  are associated with global gauge transformations, all the other soft charges we discussed here correspond to LGTs which are singular either at south or north pole of the celestial sphere. In particular, these LGTs are meromorphic (locally holomorphic) functions in the Poincar\'e  coordinates of the sphere. 
The electric and magnetic charges associated with such singular LGTs form  infinite copies of Heisenberg algebra \eqref{algebra-spatial}. The flow generated  by these LGTs on the phase space has been depicted in figure \ref{PSflow}.

For the physical interpretation of this result consider e.g. ${P}^E, Q^B_0$ which do not commute.  ${P}^E$ is  generator of boundary gauge transformation $A \rightarrow A+ d \ln{z}$ and $Q^B_0$  measures the total magnetic monopole  charge. The difference between the reference solution $A=0$ and $A=d \ln{z}$ which is generated by the mentioned gauge transformation can be attributed to  a Dirac string  piercing the celestial sphere at north and south poles. The $(2+1)$ dimensional observer at the boundary who uses the coordinates $z, \bz$ has only access to one of the two intersection points and sees this as a boundary magnetic monopole, as depicted in figure \ref{CSphere}. This interpretation can be extended to the other conjugate pairs in the set of generators. For instance $Q^E_{-1}$  generates the gauge transformation $A \rightarrow A+ d (\frac{1}{z})$. Since $\frac{1}{z}=\lim_{\epsilon\to 0}(\ln(z+\epsilon)-\ln z)/\epsilon$, this LGT generates two opposite sign magnetic monopoles with magnitude ${1}/{\epsilon}$ at a separation $\epsilon$. This is nothing but a magnetic dipole at $z=0$ at the boundary. Therefore, $Q^E_{-1}$ generates a boundary magnetic dipole, or equivalently two  Dirac strings of opposite orientation. In general, the electric charges $Q_{-n}^E, n>0$ generate boundary magnetic $2^{n}$-poles, while the magnetic charges $Q_{-n}^B$ generate boundary electric multipoles. 
\begin{figure}[t]
    \centering
    \captionsetup{width=.8\linewidth}
    \includegraphics[scale=.4]{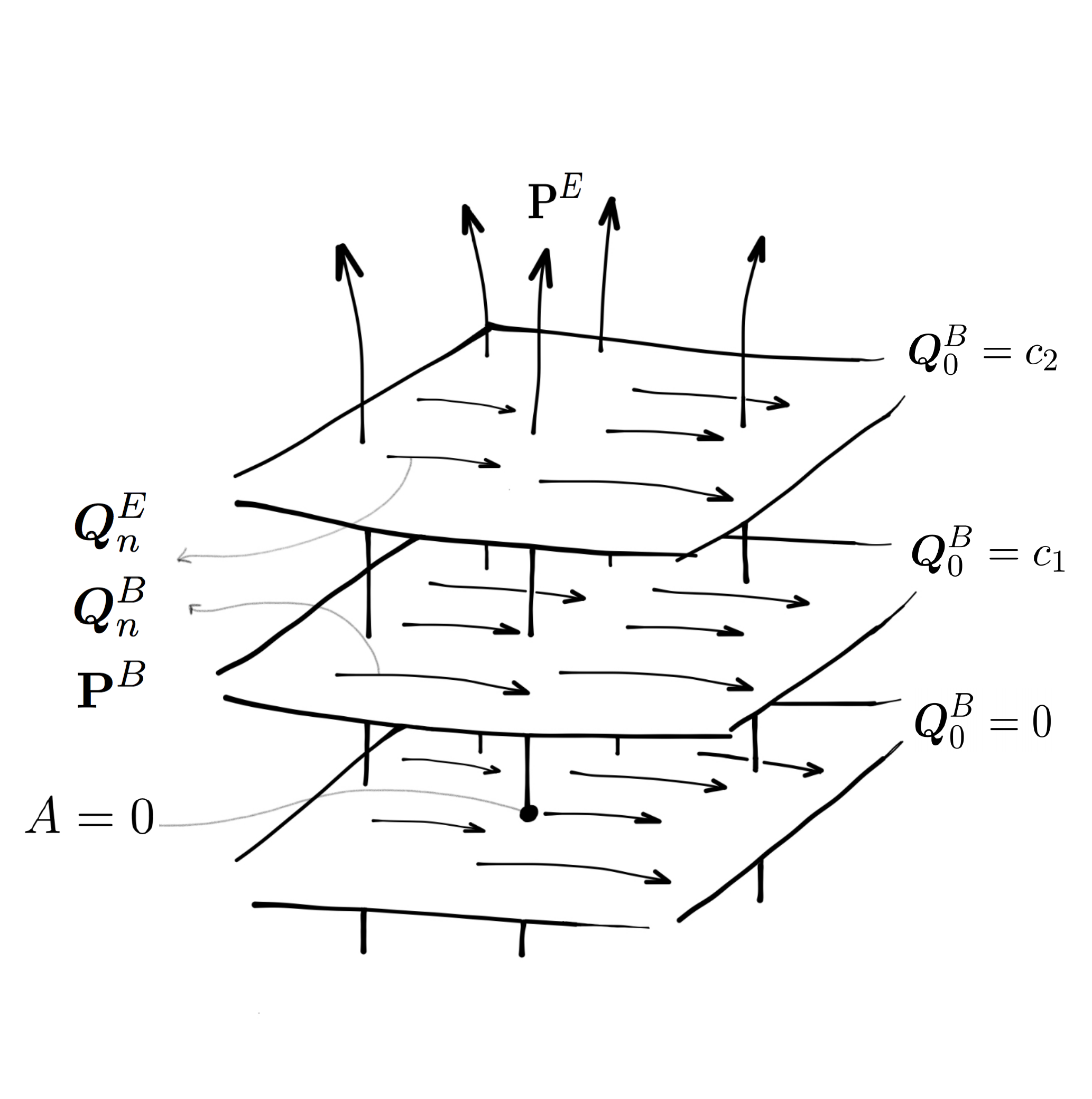}
    \caption{The phase space of soft charges and a depiction of \eqref{algebra-spatial}. The horizontal planes depict configurations of given constant magnetic charge. ${P}^E$  moves us between these horizontal planes while  $Q^E_n$, $Q^B_n$ and ${P}^B$  move us on each constant $Q_0^B$ plane. Vertical arrows show flows generated by ${P}^E$ and corresponds to addition of a Dirac string piercing the celestial sphere and appears as a surface magnetic charge for the local boundary observers. One could have drawn a similar figure using any other conjugate pairs of charges instead of ${P}^E$ and $Q^B_0$, e.g. ${P}^B$ and $Q^E_0$. This figure may be contrasted with the ``just electric'' residual gauge symmetry phase space which is usually considered in the Maxwell theory. In the absence of magnetic soft charges, the electric soft algebra is Abelian and  hence action of  residual electric gauge transformations does not create a flow on the phase space. }
    \label{PSflow}
\end{figure}

\begin{figure}[t]
    \centering
    \captionsetup{width=.8\linewidth}
    \includegraphics[scale=.3]{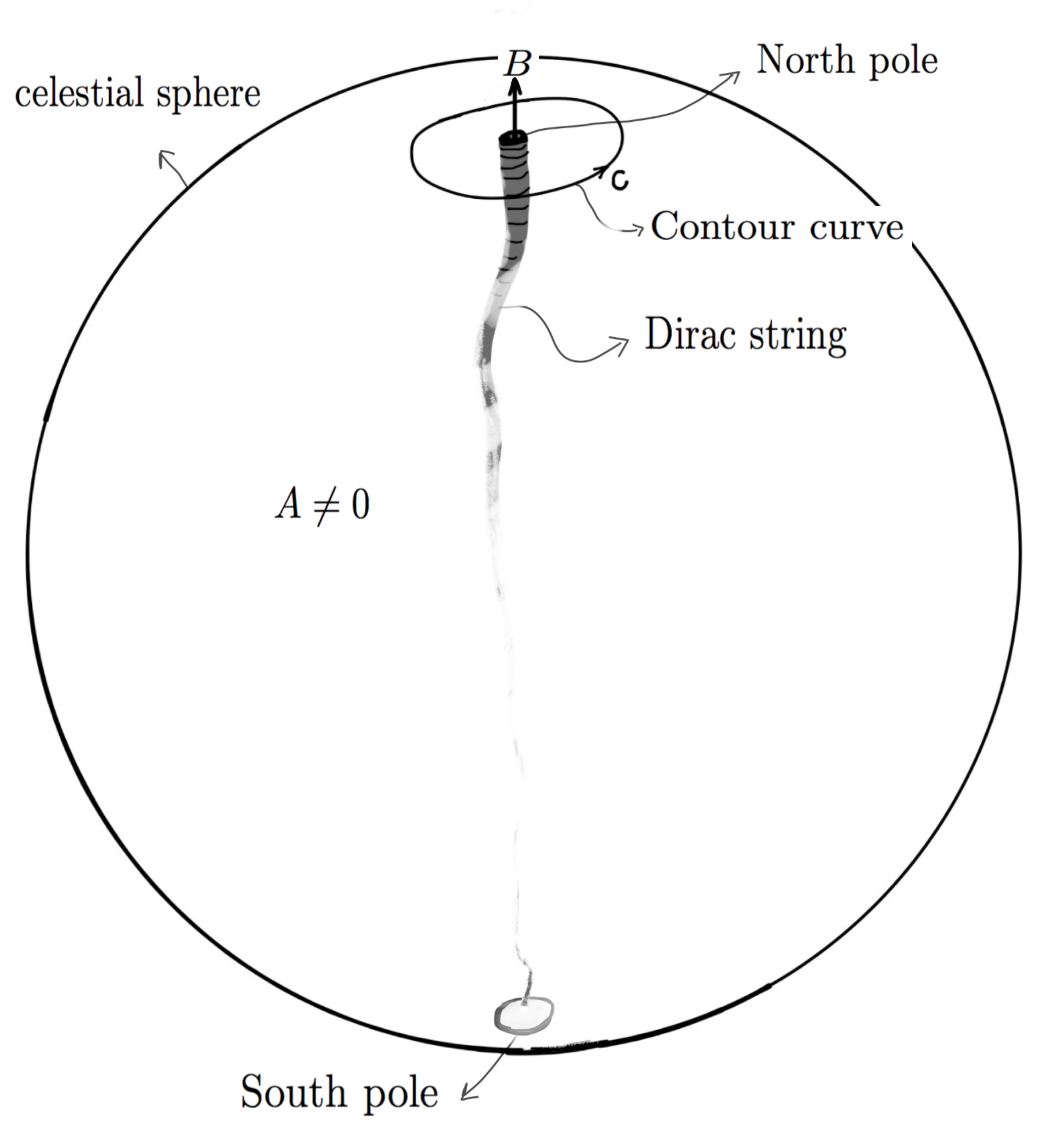}
    \caption{A depiction of the pure gauge transformation which at the celestial sphere becomes $\ln z$. This gauge transformation may be viewed as a Dirac string connecting the south and north poles of the celestial sphere with some bulk extension. It produces a ``magnetic charge'' at the north pole and an ``anti-magnetic charge'' at the south pole. The observer at the north pole who uses complex coordinates $z,\bz$ however, does not have access to the charge in the south pole and hence only sees a net magnetic charge. Similar picture may also be drawn for other higher pole charges. Note that in $2+1$ spacetime (boundary observers),  there are three residual gauge symmetry components, two for the electric one-form (or vector) and one for the magnetic two-form (or scalar). Therefore, for the boundary observer the net magnetic monopole is computed as the integral of two-form $B$ on the region confined by the contour $c$, or equivalently, $\oint_c A$.}
    \label{CSphere}
\end{figure}


\paragraph{Aharonov-Bohm phase and its generalizations as possible observables associated with soft charges.}
As we saw, ${P}^E$ generates boundary gauge transformation $A\rightarrow A+d \ln{z}$ on the boundary and can be seen as adding a Dirac string which hits the boundary at the north and south pole. We also discussed that this transformation can be measured by the boundary observer who can measure the flux of magnetic field through the enclosed part of the boundary by the contour $c$, $Q^B_0=\oint_c A$, which can be non-zero only if $c$ encircles  the singular point $z=0$. For an observer who does not have access to the singular point, this can be observed as a Ahanarov-Bohm (AB) quantum mechanical phase in a suitable quantum mechanical experiment; the AB effect provides a physical observable setup for the $A\rightarrow A+d \ln{z}$ gauge transformation; see \cite{Hamada:2017uot, Hamada:2017bgi} for further discussions. 

One can imagine a generalization of the AB phase $\Phi_{AB}=\oint_c A$ to  other singular gauge transformations associated with $Q_n, n\neq 0$ charges. In analogy with the expression of higher $n$ charges, the generalized AB factor associated with residual gauge parameter $\lambda$ may be defined as 
\be
\Phi^{\lambda}=\oint_c \lambda A\,.
\ee 
Note that although the net flux of the magnetic field is zero for higher $n$ charges, $\Phi^\lambda$ can be non-zero
for specific choice of $\lambda$ (note that $\Phi^{\lambda=1}=\Phi_{AB}$). This in principle can label some unique and well-defined quantum effects which in the case of  $\lambda=1$ is the AB effect. For example in the case of two nearby Dirac strings of opposite orientations, as discussed  above we have a ``dipole AB phase'', which could lead to its own specific quantum effect such as the effects on the quantum scattering as discussed in \cite{bogomolny2010scattering}. 

\paragraph{Memory effect and algebra of charges.} As reviewed in the introduction, memory effect is usually stated as a way to detect soft charges. On the other hand, as our analysis in this work provides an example, the information of the soft charges is fully reflected in their algebra. It is hence very desirable to provide an algebraic presentation of the memory effect. The key in our analysis is \eqref{PEMalgabra} which involves a set of ``hard charges'' (here Poincare charges $\bQ_\xi$) which do not commute with a set of ``soft charges'' (here $\bQ^E, \bQ^B$). In experiments/observations we can directly measure $\bQ_\xi$ and since they do not commute with the soft charges, a change in the soft charges yields a change in $\bQ_\xi$. The same argument can of course be made for gravitational memory effect. We intend to study develop further the algebraic statement of the memory effect.

\paragraph{More on dual symmetric theory and electric and magnetic frames.} In section \ref{DS section} we introduced a theory in which both electric and magnetic soft charges appear as residual $U(1)$ symmetries. This was achieved through a duality symmetric Maxwell theory. While we start with two electric and magnetic gauge fields $A, C$, the constraint \eqref{Dconstraints} reduces the theory to usual Maxwell on-shell. However, this leaves us with the boundary gauge transformations (LGTs) for both electric and magnetic degrees of freedom. To work out the soft charges and the associated phase space, we need to start from a (pre)symplectic structure, which is a $(3;2)$-form integrated over a Cauchy surface $\Sigma$. On the other hand the constraints \ref{Dconstraints} can be decomposed into two halves on-shell: a constraint on $\Sigma$ and the time evolution of this constraint. We should therefore only impose half of the constraints on $\Sigma$ (while the other half are guaranteed on-shell). Depending on the half we choose to impose on $\Sigma$ we end up with an electric or magnetic pictures. The expression of symplectic structure, the charges and their algebra then depend on the picture (\emph{cf} section \ref{sec: Duality charge}). These two pictures are hence to be viewed as two different basis for expanding the same set of charge and the duality charge $Q_\theta$ should be viewed as the operators rotating these two pictures (basis) into each other. As we discussed in section \ref{sec:4.4}, in the literature of duality symmetric Maxwell theory $Q_\theta$ has been dubbed as optical helicity \cite{bliokh2013dual,cameron2012electric} and the name is justified as it denotes the overall helicity (number of left polarized minus right polarized) of photons. 

The symplectic structure of the theory in electric or magnetic pictures has a bulk term (integrated over $\Sigma$) and a boundary term (integrated over the celestial sphere). The boundary term may then be viewed as the symplectic structure of a ``boundary theory'' whose degrees of freedom are labeled by the soft charges.\footnote{This boundary theory may be defined on ${\cal I}^+$ as in \cite{Cheung:2016iub, Nande:2017dba}. In this case the ``boundary theory'' will be a Euclidean 2d theory defined on the celestial sphere.} This boundary theory, as its symplectic structure indicates, resembles a Chern-Simons theory 
whose degrees of freedom are the boundary value of $A, C$ fields and $Q_\theta$ is expected to act as a symmetry on the associated phase space. This theory certainly deserves to be studied more closely along the lines of \cite{Geiller:2017xad, Geiller:2017whh, Speranza:2017gxd}.

\paragraph{Quantization of the algebra and the phase space.}

To give a semiclassical description of the theory, we should replace the Poisson bracket with Dirac brackets by replacing $\{\cdot,\cdot\}\to -i[\cdot,\cdot]$. Therefore the semiclassical version of the algebra is 
\begin{align}
[\bQ_{n}^E,\bQ_{m}^B]= 2\pi n\delta_{m+n,0}&,\qquad
[\bQ_0^E,\mathbf{P}^B]=[\bQ_0^B,\mathbf{P}^E]=-2\pi.\label{algebra-quantum-1}  
\end{align}
The algebra \eqref{algebra-quantum-1} is of the form of creation-annihilation algebra associated with a free 2d scalar theory or a one-dimensional closed string worldsheet field, consisting of a left and a right mover and $\bQ_0,\mathbf{P}$'s show its ``center of mass'' motion. To see this explicitly, let us introduce 
\be\label{alpha-QE-QB}
\begin{split}
\ball_n=\frac{1}{\sqrt{4\pi}}(\bQ_n^E+\bQ_n^B)&,\qquad
\balr_n=\frac{1}{\sqrt{4\pi}}(\bQ_{-n}^E-\bQ_{-n}^B),\cr
\bpil=\frac{-i}{\sqrt{4\pi}}(\mathbf{P}^E+\mathbf{P}^B) &,\qquad \bpir=\frac{i}{\sqrt{4\pi}}(\mathbf{P}^E-\mathbf{P}^B),
\end{split}
\ee
where one can readily check that the left and right sectors decouple and 
\begin{align}\label{alpha-algebra}
[\ball_n,\ball_m]&= [\balr_n,\balr_m]=n\delta_{m+n,0},\\
[\ball_0,\bpil]&=[\balr_0,\bpir]=i. 
\end{align}
The above algebra admits the following Hermitian conjugation:
\be
(\bQ^E_n)^\dagger=\bQ^E_{-n},\quad (\bQ^B_n)^\dagger=\bQ^B_{-n},\qquad ({\bf P}^E)^\dagger=-{\bf P}^E,\quad ({\bf P}^B)^\dagger=-{\bf P}^B,\qquad  
\ee
and hence $(\balr_n)^\dagger=\balr_{-n},\ (\ball_n)^\dagger=\ball_{-n}$ and $(\bpir)^\dagger=\bpir, (\bpil)^\dagger=\bpil$.

The ``vacuum state'' of the Hilbert space is then specified by the value of electric and magnetic charge, $|Q_0^E,Q_0^B\rangle$ such that
\be\label{vacuum}
\begin{split}
\bQ_0^E|Q_0^E,Q_0^B\rangle=Q_0^E|Q_0^E,Q_0^B\rangle,&\qquad
\bQ_0^B|Q_0^E,Q_0^B\rangle=Q_0^B|Q_0^E,Q_0^B\rangle,\cr
\ball_n|Q_0^E,Q_0^B\rangle=&\balr_n|Q_0^E,Q_0^B\rangle=0,\ n>0
\end{split}
\ee
We may take this vacuum state to have norm one. 
The ``excited states'' in the ``soft electromagnetic Hilbert space'' are then constructed by the action of $\ball_{-n}$ or $\balr_{-n},\ n>0$. However, one should note that as \eqref{alpha-QE-QB} shows, while $\ball_n, n>0$ is related to $\bQ_n$'s with $n>0$, the $\balr_n, n>0$ is related to $\bQ_{-n}$'s.  
As we will discuss below, we expect to have the following Dirac quantization condition, 
\be\label{Dirac-quantization}
Q_0^EQ_0^B=2\pi\hbar \mathbb{Z},\qquad Q_n^EQ_{-n}^B=2\pi\hbar \mathbb{Z}.
\ee
Therefore, the vacuum state and other excited states in the soft Hilbert space are expected to be specified by discrete labels.

\paragraph{Virasoro and Kac-Moody algebras from electromagnetic soft charges.} Given the two $\ball_n, \balr_n$ operators one may construct two left and right Virasoro algebras using Sugawara construction
\begin{align}
    \bLl_n=\frac12\sum_p :\ball_p \ball_{n-p}:,\qquad     \bLr_n=\frac12\sum_p :\balr_p \balr_{n-p}:
\end{align}
where : : denotes normal ordering. Each sector forms a $U(1)$ Kac-Moody algebra at central charge one
\begin{align}
\begin{split}
    [\bcL_n,\bcL_m]&=(n-m)\bcL_{n+m}+\dfrac{1}{12}(n^3-n) \delta_{n+m,0}\,,\\
    [\bcL_n,\boldsymbol{\alpha}_m]&=-m\boldsymbol{\alpha}_{m+n},\qquad 
    [\boldsymbol{\alpha}_m,\boldsymbol{\alpha}_n]= m\delta_{m+n,0},
\end{split}
\end{align}
and the left and right sectors commute with each other. Had we started with a multi-Maxwell theory with $N$ non-interacting $U(1)$ gauge fields, we would have obtained
a Virasoro of central charge $N$. Of course there is another way to obtain a Virasoro with arbitrary central charge: to add a ``twist term'' to the Virasoro generators  
$\bcL_n=\frac12\sum_p\boldsymbol{\alpha}_p\boldsymbol{\alpha}_{n-p}+i\beta n \boldsymbol{\alpha}_n$ (see e.g. \cite{Afshar:2016kjj, Afshar:2016uax, Afshar:2017okz}) to obtain a Virasoro at central charge $c=1+12\beta^2$. This twist term is behaving like a linear dilaton background. In this twisted construction, however, $\boldsymbol{\alpha}_n$'s do not remain as $U(1)$ current of weight one, $[\bcL,\boldsymbol{\alpha}]$ commutator will have an anomaly term \cite{Afshar:2017okz}.

One can show that the combination $(\bLl_0-\bLr_0)$ generates the same algebra with soft charges as the spin $\mathcal{J}$ as in \eqref{S-QE-QB}.\footnote{One should note that here we are working in the electric picture and consistently dropping the magnetic picture charges, $\tilde Q$'s.}  After expanding in electric and magnetic charges, we find
\begin{align}
   \bLl_0-\bLr_0=\frac{1}{2\pi}\sum_{p\in \mathbb{Z}}\ :\bQ^E_{-p}\bQ^B_p: 
\end{align}
As it is related to the spin operator, the spectrum of $\bLl_0-\bLr_0$ is expected to be quantized, as in \eqref{Dirac-quantization}; in our construction the Bohr-type quantization of the angular momentum gives rise to the Dirac quantization of electric and magnetic charges.
We comment that, 
\begin{align}
   [\boldsymbol{{\cal L}}_0, \bQ_n^E]=-n \bQ^B_n,\quad    [\boldsymbol{{\cal L}}_0, \bQ_n^B]=+n \bQ^E_n, \qquad 
   \boldsymbol{{\cal L}}_0\equiv \bLl_0+\bLr_0.
\end{align}
The above shows that $\boldsymbol{{\cal L}}_0$ is different from the duality charge operator $\bQ_\theta$; the latter is more like a number operator which counts the difference between number of left and right helicities, as discussed above. 

\paragraph{Extension to higher forms.} The analysis of this paper can be extended to $(p+1)$-form theories in $2p+4$ dimensions. The electric soft charges of such form theories was carried out in \cite{Afshar:2018apx} where it was shown that for generic $p>0$ cases the residual gauge symmetry charges appear in three classes, one of which, the ``exact charges'' in the terminology of \cite{Afshar:2018apx}, has no counterpart in the Maxwell theory. These exact charges satisfy a non-commuting algebra. We expect our result for the Maxwell case, that the electric and magnetic charges are non-commuting, extends to these higher form cases. Therefore, we expect there are two classes of non-commuting soft charges for $p>0$ cases. It is desirable to verify this expectation and study its physical implications. 

\subsection*{Acknowledgement}
We would like to thank Hamid Afshar, Sajad Aghapour, Beatrice Bonga, Miguel Campiglia, Erfan Esmaili, Marc Geiller, Daniel Grumiller, Ghadir Jafari, Gary Shiu and Andy Strominger for fruitful discussions or comments. AS would like to thank Laurent Freidel for motivating discussions during his visit to Perimeter institute. AS would like to thank Geoffrey Compere and the ERC Starting Grant 335146 for a visit at ULB during the last stages of this work. 
This work is supported in part by Iranian National Science Foundation (INSF) junior research chair in black hole physics, grant no. 950124 and ICTP network scheme NT-04 funds.

\appendix
\section{Contour integrals}\label{App:A}

In our analysis of the charge algebra we need to compute the following integrals:
\begin{align}
I=\int (d\alpha_n)\wedge (d \beta_m)=2 \pi i m  \bigg[(\alpha_n \beta_m-\bar{\alpha}_n \bar{\beta}_m)\delta_{m+n,0}+\lim_{r_c \rightarrow 0} r_c^{2m} (\bar{\alpha}_n \beta_m- \alpha_n\bar{\beta}_m)\delta_{m,n}\bigg],
\end{align}
with,
\begin{align}
\alpha_n=\alpha_n z^n+ \bar{\alpha}_n \bar{z}^n, \qquad 
\beta_n=\beta_n z^n+ \bar{\beta}_n \bar{z}^n.
\end{align}
To compute $I$ we have used the formulas,
\begin{align}
\int_\mathds{C} dz^n \wedge dz^m&=2\pi i m \delta_{m+n,0},\\
\int_\mathds{C} d\bar{z}^n \wedge dz^m&=\lim_{r_c \rightarrow \infty} r_c^{2m} 2\pi i m \delta_{m,n},
\end{align}
where $r_c$ is the radius of the contour around poles.  

One can check the first and the second formula by directly computing the integral of surface element $dz^n \wedge dz^m$ or by using the Stokes theorem and turn it to a contour integral around poles. The second method is as follows,
\begin{align}
\int_\mathds{C} dz^n \wedge dz^m=\int_{\mathds{U}_0} dz^n \wedge dz^m=
\int_{\mathds{U}_0} d[z^n dz^m]=\int_{\partial\mathds{U}_0} z^n dz^m=
\int_{\partial\mathds{U}_0} m z^{m+n-1} dz=2\pi i m \delta_{m+n,0}
\end{align}  
where $\mathds{U}_0$ is a region of $\mathds{C}$ while $\{0\} \in \mathds{U}_0$.
Note that $dz^n \wedge dz^m$ is a zero two-form on $\mathds{C}-\{0\}$ and has 
singularity at $\{0\}$. So the integral of this two-form over any region except $\mathds{U}_0$
is zero.
The other way to calculate this integral is using the identities,
\begin{align}
\partial_{\bz}(z^{-m})&=\frac{2 \pi i}{(-1)^{m-1}(m-1)!}\partial_{z}^{m-1}\delta^2(z), \qquad \qquad & m >0\\
\partial_{\bz}(z^{-m})&=0,   & m\leq 0\\
z^m \partial_{z}^n \delta(z)&=(-1)^n n! \delta(z) \delta_{m,n}\,, \qquad \qquad &m >0 \\
z^{-m} \partial_{z}^n \delta(z)&=\frac{(m+n-1)!}{(m-1)!} z^{-(m+n)} \delta(z),  \qquad \qquad &m >0.
\end{align}
Note that when both $m$ and $n$ are positive or negative, the 
integral is zero. So, we assume without losing of generality that $n <0$ and $m>0$.
\begin{align}
\int_\mathds{C} dz^n \wedge dz^m&=\int_\mathds{C}m\partial_{\bz}(z^{n}) z^{m-1} d\bar{z} \wedge dz=\int_\mathds{C}\frac{2 m \pi i z^{m-1}}{(-1)^{-n-1}(-n-1)!}\partial_{z}^{-n-1}\delta^2(z)d\bar{z} \wedge dz\\
&= \int_\mathds{C}\frac{2 m \pi i (-1)^{-n-1}(-n-1)!}{(-1)^{-n-1}(-n-1)!}\delta^2(z) \delta_{m+n,0}d\bar{z} \wedge dz
=2\pi i m \delta_{m+n,0}
\end{align}  
for the case $n>0, m>0$ the integral is manifestly zero. But for the case $n<0, m<0$, we have,
\begin{align}
\int_\mathds{C} dz^n \wedge dz^m&=\int_\mathds{C} (\partial_{z} z^n \partial_{\bz} z^m-\partial_{\bz} z^n \partial_{z} z^m) dz\wedge d \bar{z}=\int_\mathds{C} 2 \pi i [-(n+m+1)]! \delta^2(z) z^{(n+m)}\\& \times\Big[\frac{1}{(-1)^{-(m+1)}(-m-1)!(-n)!}-\frac{1}{(-1)^{-(n+1)}(-n-1)!(-m)!}\big]dz\wedge d \bar{z}=0
\end{align}

\section{An alternative formulation: Complexified Maxwell theory}\label{App:B}

In this section, we show that the construction above arise in a natural way in the complexification of Maxwell theory \cite{bliokh2013dual}. Assume that dynamical gauge field is complex instead of real, i.e. define the complex gauge field as $\cA=A+iC$. The complex field strength is defined as $\cF=d\cA=F+iG$; $\cF$ is a two-form under usual Lorentz transformations. The constraint $G=\ast F$ is demonstrated here as 
\begin{align}
    \ast \cF=-i \cF,
\end{align}
which is like the self-duality condition in ordinary Euclidean Maxwell theory. The Lagrangian can be written as 
\begin{align}
    \cL=-\dfrac{1}{2}\cF\wedge\ast \bar \cF,
\end{align}
 and the equations of motion are
\begin{align}
    \partial_\mu \cF^{\mu\nu}=0,\qquad \partial_\mu \bar \cF^{\mu\nu}=0.
\end{align}
The theory is invariant under the gauge transformation
\begin{align}
    \cA\to \cA+d{\Lambda},
\end{align}
where $\Lambda=f+ig$ and  $\bar\Lambda=f-ig$ is the complex conjugate of $\Lambda$. {The duality symmetry transformation \eqref{Duality-rotation-A,C} appears as a global $U(1)$ symmetry $\cA\to e^{-i\theta}\cA$. This global symmetry cannot be gauged \cite{Bunster:2010wv}. }

Now one may compute the Noether charges associated to these gauge symmetries. One finds 
\begin{align}
    \bcQ_\Lambda=-\dfrac{1}{2}\oint d\Sigma_{\mu\nu}(\Lambda \bar \cF^{\mu\nu}+\bar\Lambda \cF^{\mu\nu})
    =\oint (\Lambda \,\partial_z A_{\bar z} +\bar \Lambda\, \partial_{\bar z} A_z).
\end{align}
Note that the charge is a real function. One may alternatively compute  the electric and magnetic charges at null infinity:
\begin{align}
    \cQ_\Lambda&=Q^E_f+Q^B_g.
\end{align}
The above is of course compatible with \eqref{ECharge} and \eqref{BCharge}. 

\paragraph{Algebra of all charges.}

Given two complex gauge variables $\Lambda=f+ig$ and $\tilde{\Lambda}=\tilde f+i\tilde g$, we can compute the algebra of charges, e.g. in spatial slicing using the analysis of section \ref{sec:2},
\begin{align}\label{charge-algebra-complex}
    \{\bcQ_\Lambda,\bcQ_{\tilde\Lambda}\}&=\{\bQ^E_f,\bQ^B_{\tilde g}\}-\{\bQ^E_{\tilde f},\bQ^B_g\}=\oint (df\wedge d\tilde g-d\tilde f \wedge dg)\\
    &=\mathrm{Im}\oint d\Lambda\wedge d\tilde \Lambda
\end{align}
The above for pure real or pure imaginary $\Lambda$ reproduces \eqref{AlgebraNull}.
Algebra of charges at null infinity may also be worked out along the lines of section \ref{Null-section}. This yields \eqref{charge-algebra-complex} which may be shown to be exactly the same as \eqref{AlgebraNull}.


\bibliographystyle{JHEP} 
\bibliography{mybib}{}

\providecommand{\href}[2]{#2}\begingroup\raggedright\begin{thebibliography}{10}

\bibitem{Strominger:2017zoo}
A.~Strominger, \emph{{Lectures on the Infrared Structure of Gravity and Gauge
  Theory}},  \href{https://arxiv.org/abs/1703.05448}{{\ttfamily 1703.05448}}.

\bibitem{Lee:1990nz}
J.~Lee and R.~M. Wald, \emph{{Local symmetries and constraints}},
  \href{https://doi.org/10.1063/1.528801}{\emph{J. Math. Phys.} {\bfseries 31}
  (1990) 725--743}.

\bibitem{Ashtekar:1990gc}
A.~Ashtekar, L.~Bombelli and O.~Reula, \emph{{The Covariant Phase Space Of
  Asymptotically Flat Gravitational Fields}}, {\emph{PRINT-90-0318 (SYRACUSE)}
  (1990) }.

\bibitem{Ashtekar:1987tt}
A.~Ashtekar, \emph{{Asympotitc Quantization: Based on 1984 Naples Lectures}},
  {\emph{Naples, Italy: Bibliopolis (1987) 107 P. (Monographs and Textbooks in
  physical science, 2)} (1987) }.

\bibitem{Henneaux:1985kr}
M.~Henneaux, \emph{{Hamiltonian Form of the Path Integral for Theories with a
  Gauge Freedom}},
  \href{https://doi.org/10.1016/0370-1573(85)90103-6}{\emph{Phys. Rept.}
  {\bfseries 126} (1985) 1--66}.

\bibitem{Henneaux:1992ig}
M.~Henneaux and C.~Teitelboim, \emph{{Quantization of gauge systems}}.
\newblock Princeton, USA: Univ. Pr. (1992) 520 p, 1992.

\bibitem{Brown:1986nw}
J.~D. Brown and M.~Henneaux, \emph{{Central Charges in the Canonical
  Realization of Asymptotic Symmetries: An Example from Three-Dimensional
  Gravity}}, \href{https://doi.org/10.1007/BF01211590}{\emph{Commun. Math.
  Phys.} {\bfseries 104} (1986) 207--226}.

\bibitem{Henneaux:2018gfi}
M.~Henneaux and C.~Troessaert, \emph{{Asymptotic symmetries of electromagnetism
  at spatial infinity}},
  \href{https://doi.org/10.1007/JHEP05(2018)137}{\emph{JHEP} {\bfseries 05}
  (2018) 137}, [\href{https://arxiv.org/abs/1803.10194}{{\ttfamily
  1803.10194}}].

\bibitem{Henneaux:2018hdj}
M.~Henneaux and C.~Troessaert, \emph{{Hamiltonian structure and asymptotic
  symmetries of the Einstein-Maxwell system at spatial infinity}},
  \href{https://arxiv.org/abs/1805.11288}{{\ttfamily 1805.11288}}.

\bibitem{Kapec:2015ena}
D.~Kapec, M.~Pate and A.~Strominger, \emph{{New Symmetries of QED}},
  \href{https://arxiv.org/abs/1506.02906}{{\ttfamily 1506.02906}}.

\bibitem{Bieri:2011zb}
L.~Bieri, P.~Chen and S.-T. Yau, \emph{{The Electromagnetic Christodoulou
  Memory Effect and its Application to Neutron Star Binary Mergers}},
  \href{https://doi.org/10.1088/0264-9381/29/21/215003}{\emph{Class. Quant.
  Grav.} {\bfseries 29} (2012) 215003},
  [\href{https://arxiv.org/abs/1110.0410}{{\ttfamily 1110.0410}}].

\bibitem{Bieri:2013hqa}
L.~Bieri and D.~Garfinkle, \emph{{An electromagnetic analogue of gravitational
  wave memory}},
  \href{https://doi.org/10.1088/0264-9381/30/19/195009}{\emph{Class. Quant.
  Grav.} {\bfseries 30} (2013) 195009},
  [\href{https://arxiv.org/abs/1307.5098}{{\ttfamily 1307.5098}}].

\bibitem{Susskind:2015hpa}
L.~Susskind, \emph{{Electromagnetic Memory}},
  \href{https://arxiv.org/abs/1507.02584}{{\ttfamily 1507.02584}}.

\bibitem{Pasterski:2015zua}
S.~Pasterski, \emph{{Asymptotic Symmetries and Electromagnetic Memory}},
  \href{https://doi.org/10.1007/JHEP09(2017)154}{\emph{JHEP} {\bfseries 09}
  (2017) 154}, [\href{https://arxiv.org/abs/1505.00716}{{\ttfamily
  1505.00716}}].

\bibitem{Hamada:2017uot}
Y.~Hamada, M.-S. Seo and G.~Shiu, \emph{{Large gauge transformations and little
  group for soft photons}},
  \href{https://doi.org/10.1103/PhysRevD.96.105013}{\emph{Phys. Rev.}
  {\bfseries D96} (2017) 105013},
  [\href{https://arxiv.org/abs/1704.08773}{{\ttfamily 1704.08773}}].

\bibitem{Hamada:2018vrw}
Y.~Hamada and G.~Shiu, \emph{{Infinite Set of Soft Theorems in Gauge-Gravity
  Theories as Ward-Takahashi Identities}},
  \href{https://doi.org/10.1103/PhysRevLett.120.201601}{\emph{Phys. Rev. Lett.}
  {\bfseries 120} (2018) 201601},
  [\href{https://arxiv.org/abs/1801.05528}{{\ttfamily 1801.05528}}].

\bibitem{Hirai:2018ijc}
H.~Hirai and S.~Sugishita, \emph{{Conservation Laws from Asymptotic Symmetry
  and Subleading Charges in QED}},
  \href{https://arxiv.org/abs/1805.05651}{{\ttfamily 1805.05651}}.

\bibitem{Flanagan:2014kfa}
E.~E. Flanagan and D.~A. Nichols, \emph{{Observer dependence of angular
  momentum in general relativity and its relationship to the gravitational-wave
  memory effect}}, \href{https://doi.org/10.1103/PhysRevD.92.084057,
  10.1103/PhysRevD.93.049905}{\emph{Phys. Rev.} {\bfseries D92} (2015) 084057},
  [\href{https://arxiv.org/abs/1411.4599}{{\ttfamily 1411.4599}}].

\bibitem{Bieri:2015yia}
L.~Bieri, D.~Garfinkle and S.-T. Yau, \emph{{Gravitational Waves and Their
  Memory in General Relativity}},
  \href{https://arxiv.org/abs/1505.05213}{{\ttfamily 1505.05213}}.

\bibitem{Tolish:2014oda}
A.~Tolish, L.~Bieri, D.~Garfinkle and R.~M. Wald, \emph{{Examination of a
  simple example of gravitational wave memory}},
  \href{https://doi.org/10.1103/PhysRevD.90.044060}{\emph{Phys. Rev.}
  {\bfseries D90} (2014) 044060},
  [\href{https://arxiv.org/abs/1405.6396}{{\ttfamily 1405.6396}}].

\bibitem{Pate:2017fgt}
M.~Pate, A.-M. Raclariu and A.~Strominger, \emph{{Gravitational Memory in
  Higher Dimensions}},  \href{https://arxiv.org/abs/1712.01204}{{\ttfamily
  1712.01204}}.

\bibitem{Giddings:2018umg}
S.~B. Giddings and A.~Kinsella, \emph{{Gauge-invariant observables,
  gravitational dressings, and holography in AdS}},
  \href{https://arxiv.org/abs/1802.01602}{{\ttfamily 1802.01602}}.

\bibitem{Kulish:1970ut}
P.~P. Kulish and L.~D. Faddeev, \emph{{Asymptotic conditions and infrared
  divergences in quantum electrodynamics}},
  \href{https://doi.org/10.1007/BF01066485}{\emph{Theor. Math. Phys.}
  {\bfseries 4} (1970) 745}.

\bibitem{Gabai:2016kuf}
B.~Gabai and A.~Sever, \emph{{Large gauge symmetries and asymptotic states in
  QED}}, \href{https://doi.org/10.1007/JHEP12(2016)095}{\emph{JHEP} {\bfseries
  12} (2016) 095}, [\href{https://arxiv.org/abs/1607.08599}{{\ttfamily
  1607.08599}}].

\bibitem{Herdegen:2016bio}
A.~Herdegen, \emph{{Asymptotic structure of electrodynamics revisited}},
  \href{https://doi.org/10.1007/s11005-017-0948-9}{\emph{Lett. Math. Phys.}
  {\bfseries 107} (2017) 1439--1470},
  [\href{https://arxiv.org/abs/1604.04170}{{\ttfamily 1604.04170}}].

\bibitem{Montonen:1977sn}
C.~Montonen and D.~I. Olive, \emph{{Magnetic Monopoles as Gauge Particles?}},
  \href{https://doi.org/10.1016/0370-2693(77)90076-4}{\emph{Phys. Lett.}
  {\bfseries 72B} (1977) 117--120}.

\bibitem{Seiberg:1994rs}
N.~Seiberg and E.~Witten, \emph{{Electric - magnetic duality, monopole
  condensation, and confinement in N=2 supersymmetric Yang-Mills theory}},
  \href{https://doi.org/10.1016/0550-3213(94)90124-4,
  10.1016/0550-3213(94)00449-8}{\emph{Nucl. Phys.} {\bfseries B426} (1994)
  19--52}, [\href{https://arxiv.org/abs/hep-th/9407087}{{\ttfamily
  hep-th/9407087}}].

\bibitem{Strominger:2015bla}
A.~Strominger, \emph{{Magnetic Corrections to the Soft Photon Theorem}},
  \href{https://doi.org/10.1103/PhysRevLett.116.031602}{\emph{Phys. Rev. Lett.}
  {\bfseries 116} (2016) 031602},
  [\href{https://arxiv.org/abs/1509.00543}{{\ttfamily 1509.00543}}].

\bibitem{Campiglia:2016hvg}
M.~Campiglia and A.~Laddha, \emph{{Subleading soft photons and large gauge
  transformations}}, \href{https://doi.org/10.1007/JHEP11(2016)012}{\emph{JHEP}
  {\bfseries 11} (2016) 012},
  [\href{https://arxiv.org/abs/1605.09677}{{\ttfamily 1605.09677}}].

\bibitem{Zwanziger:1968rs}
D.~Zwanziger, \emph{{Quantum field theory of particles with both electric and
  magnetic charges}},
  \href{https://doi.org/10.1103/PhysRev.176.1489}{\emph{Phys. Rev.} {\bfseries
  176} (1968) 1489--1495}.

\bibitem{Zwanziger:1970hk}
D.~Zwanziger, \emph{{Local Lagrangian quantum field theory of electric and
  magnetic charges}}, \href{https://doi.org/10.1103/PhysRevD.3.880}{\emph{Phys.
  Rev.} {\bfseries D3} (1971) 880}.

\bibitem{Cardona:2015woa}
C.~Cardona, \emph{{Asymptotic Symmetries of Yang-Mills with Theta Term and
  Monopoles}},  \href{https://arxiv.org/abs/1504.05542}{{\ttfamily
  1504.05542}}.

\bibitem{Freidel:2018fsk}
L.~Freidel and D.~Pranzetti, \emph{{Electromagnetic duality and central
  charge}},  \href{https://arxiv.org/abs/1806.03161}{{\ttfamily 1806.03161}}.

\bibitem{Donnelly:2016auv}
W.~Donnelly and L.~Freidel, \emph{{Local subsystems in gauge theory and
  gravity}}, \href{https://doi.org/10.1007/JHEP09(2016)102}{\emph{JHEP}
  {\bfseries 09} (2016) 102},
  [\href{https://arxiv.org/abs/1601.04744}{{\ttfamily 1601.04744}}].

\bibitem{Seraj:2016cym}
A.~Seraj, \emph{{Conserved charges, surface degrees of freedom, and black hole
  entropy}}, Ph.D. thesis, IPM, Tehran, 2016.
\newblock \href{https://arxiv.org/abs/1603.02442}{{\ttfamily 1603.02442}}.

\bibitem{Compere:2018aar}
A.~Fiorucci and G.~Comp\`ere, \emph{{Advanced Lectures in General Relativity}},
  Ph.D. thesis, Brussels U., PTM, 2018.
\newblock \href{https://arxiv.org/abs/1801.07064}{{\ttfamily 1801.07064}}.

\bibitem{Woodhouse:1980pa}
N.~Woodhouse, \emph{{Geometric Quantization}}, {\emph{Oxford, Uk: Clarendon
  (1980) 316 P. (Oxford Mathematical Monographs)} (1980) }.

\bibitem{Jackson:1998nia}
J.~D. Jackson, \emph{{Classical Electrodynamics}}.
\newblock Wiley, 1998.

\bibitem{Seraj:2016jxi}
A.~Seraj, \emph{{Multipole charge conservation and implications on
  electromagnetic radiation}},
  \href{https://doi.org/10.1007/JHEP06(2017)080}{\emph{JHEP} {\bfseries 06}
  (2017) 080}, [\href{https://arxiv.org/abs/1610.02870}{{\ttfamily
  1610.02870}}].

\bibitem{Campiglia:2015qka}
M.~Campiglia and A.~Laddha, \emph{{Asymptotic symmetries of QED and
  Weinberg’s soft photon theorem}},
  \href{https://doi.org/10.1007/JHEP07(2015)115}{\emph{JHEP} {\bfseries 07}
  (2015) 115}, [\href{https://arxiv.org/abs/1505.05346}{{\ttfamily
  1505.05346}}].

\bibitem{Geiller:2017xad}
M.~Geiller, \emph{{Edge modes and corner ambiguities in 3d Chern–Simons
  theory and gravity}},
  \href{https://doi.org/10.1016/j.nuclphysb.2017.09.010}{\emph{Nucl. Phys.}
  {\bfseries B924} (2017) 312--365},
  [\href{https://arxiv.org/abs/1703.04748}{{\ttfamily 1703.04748}}].

\bibitem{Geiller:2017whh}
M.~Geiller, \emph{{Lorentz-diffeomorphism edge modes in 3d gravity}},
  \href{https://doi.org/10.1007/JHEP02(2018)029}{\emph{JHEP} {\bfseries 02}
  (2018) 029}, [\href{https://arxiv.org/abs/1712.05269}{{\ttfamily
  1712.05269}}].

\bibitem{Speranza:2017gxd}
A.~J. Speranza, \emph{{Local phase space and edge modes for
  diffeomorphism-invariant theories}},
  \href{https://doi.org/10.1007/JHEP02(2018)021}{\emph{JHEP} {\bfseries 02}
  (2018) 021}, [\href{https://arxiv.org/abs/1706.05061}{{\ttfamily
  1706.05061}}].

\bibitem{He:2014cra}
T.~He, P.~Mitra, A.~P. Porfyriadis and A.~Strominger, \emph{{New Symmetries of
  Massless QED}}, \href{https://doi.org/10.1007/JHEP10(2014)112}{\emph{JHEP}
  {\bfseries 10} (2014) 112},
  [\href{https://arxiv.org/abs/1407.3789}{{\ttfamily 1407.3789}}].

\bibitem{Wald:1993nt}
R.~M. Wald, \emph{{Black hole entropy is the Noether charge}},
  \href{https://doi.org/10.1103/PhysRevD.48.R3427}{\emph{Phys. Rev.} {\bfseries
  D48} (1993) R3427--R3431},
  [\href{https://arxiv.org/abs/gr-qc/9307038}{{\ttfamily gr-qc/9307038}}].

\bibitem{Iyer:1994ys}
V.~Iyer and R.~M. Wald, \emph{{Some properties of Noether charge and a proposal
  for dynamical black hole entropy}},
  \href{https://doi.org/10.1103/PhysRevD.50.846}{\emph{Phys. Rev.} {\bfseries
  D50} (1994) 846--864}, [\href{https://arxiv.org/abs/gr-qc/9403028}{{\ttfamily
  gr-qc/9403028}}].

\bibitem{bliokh2013dual}
K.~Y. Bliokh, A.~Y. Bekshaev and F.~Nori, \emph{Dual electromagnetism:
  helicity, spin, momentum and angular momentum}, {\emph{New Journal of
  Physics} {\bfseries 15} (2013) 033026}.

\bibitem{cameron2012electric}
R.~P. Cameron and S.~M. Barnett, \emph{Electric--magnetic symmetry and
  noether's theorem}, {\emph{New Journal of Physics} {\bfseries 14} (2012)
  123019}.

\bibitem{Deser:1976iy}
S.~Deser and C.~Teitelboim, \emph{{Duality Transformations of Abelian and
  Nonabelian Gauge Fields}},
  \href{https://doi.org/10.1103/PhysRevD.13.1592}{\emph{Phys. Rev.} {\bfseries
  D13} (1976) 1592--1597}.

\bibitem{Barnich:2007uu}
G.~Barnich and A.~Gomberoff, \emph{{Dyons with potentials: Duality and black
  hole thermodynamics}},
  \href{https://doi.org/10.1103/PhysRevD.78.025025}{\emph{Phys. Rev.}
  {\bfseries D78} (2008) 025025},
  [\href{https://arxiv.org/abs/0705.0632}{{\ttfamily 0705.0632}}].

\bibitem{Bunster:2018yjr}
C.~Bunster, A.~Gomberoff and A.~Pérez, \emph{{Regge-Teitelboim analysis of the
  symmetries of electromagnetic and gravitational fields on asymptotically null
  spacelike surfaces}},  \href{https://arxiv.org/abs/1805.03728}{{\ttfamily
  1805.03728}}.

\bibitem{Regge:1974zd}
T.~Regge and C.~Teitelboim, \emph{{Role of Surface Integrals in the Hamiltonian
  Formulation of General Relativity}},
  \href{https://doi.org/10.1016/0003-4916(74)90404-7}{\emph{Annals Phys.}
  {\bfseries 88} (1974) 286}.

\bibitem{Bhattacharyya:2017obx}
A.~Bhattacharyya, L.-Y. Hung and Y.~Jiang, \emph{{Null hypersurface
  quantization, electromagnetic duality and asympotic symmetries of Maxwell
  theory}}, \href{https://doi.org/10.1007/JHEP03(2018)027}{\emph{JHEP}
  {\bfseries 03} (2018) 027},
  [\href{https://arxiv.org/abs/1708.05606}{{\ttfamily 1708.05606}}].

\bibitem{Hamada:2017bgi}
Y.~Hamada, M.-S. Seo and G.~Shiu, \emph{{Electromagnetic Duality and the
  Electric Memory Effect}},
  \href{https://doi.org/10.1007/JHEP02(2018)046}{\emph{JHEP} {\bfseries 02}
  (2018) 046}, [\href{https://arxiv.org/abs/1711.09968}{{\ttfamily
  1711.09968}}].

\bibitem{iyer1994some}
V.~Iyer and R.~M. Wald, \emph{Some properties of the noether charge and a
  proposal for dynamical black hole entropy}, {\emph{Physical review D}
  {\bfseries 50} (1994) 846}.

\bibitem{cameron2012optical}
R.~P. Cameron, S.~M. Barnett and A.~M. Yao, \emph{Optical helicity, optical
  spin and related quantities in electromagnetic theory}, {\emph{New Journal of
  Physics} {\bfseries 14} (2012) 053050}.

\bibitem{Weinberg:1964ew}
S.~Weinberg, \emph{{Photons and Gravitons in s Matrix Theory: Derivation of
  Charge Conservation and Equality of Gravitational and Inertial Mass}},
  \href{https://doi.org/10.1103/PhysRev.135.B1049}{\emph{Phys. Rev.} {\bfseries
  135} (1964) B1049--B1056}.

\bibitem{Barnich:2010eb}
G.~Barnich and C.~Troessaert, \emph{{Aspects of the BMS/CFT correspondence}},
  \href{https://doi.org/10.1007/JHEP05(2010)062}{\emph{JHEP} {\bfseries 05}
  (2010) 062}, [\href{https://arxiv.org/abs/1001.1541}{{\ttfamily 1001.1541}}].

\bibitem{Barnich:2011mi}
G.~Barnich and C.~Troessaert, \emph{{BMS charge algebra}},
  \href{https://doi.org/10.1007/JHEP12(2011)105}{\emph{JHEP} {\bfseries 12}
  (2011) 105}, [\href{https://arxiv.org/abs/1106.0213}{{\ttfamily 1106.0213}}].

\bibitem{Barnich:2017ubf}
G.~Barnich, \emph{{Centrally extended BMS4 Lie algebroid}},
  \href{https://doi.org/10.1007/JHEP06(2017)007}{\emph{JHEP} {\bfseries 06}
  (2017) 007}, [\href{https://arxiv.org/abs/1703.08704}{{\ttfamily
  1703.08704}}].

\bibitem{bogomolny2010scattering}
E.~Bogomolny, S.~Mashkevich and S.~Ouvry, \emph{Scattering on two
  aharonov--bohm vortices with opposite fluxes}, {\emph{Journal of Physics A:
  Mathematical and Theoretical} {\bfseries 43} (2010) 354029}.

\bibitem{Cheung:2016iub}
C.~Cheung, A.~de~la Fuente and R.~Sundrum, \emph{{4D scattering amplitudes and
  asymptotic symmetries from 2D CFT}},
  \href{https://doi.org/10.1007/JHEP01(2017)112}{\emph{JHEP} {\bfseries 01}
  (2017) 112}, [\href{https://arxiv.org/abs/1609.00732}{{\ttfamily
  1609.00732}}].

\bibitem{Nande:2017dba}
A.~Nande, M.~Pate and A.~Strominger, \emph{{Soft Factorization in QED from 2D
  Kac-Moody Symmetry}},
  \href{https://doi.org/10.1007/JHEP02(2018)079}{\emph{JHEP} {\bfseries 02}
  (2018) 079}, [\href{https://arxiv.org/abs/1705.00608}{{\ttfamily
  1705.00608}}].

\bibitem{Afshar:2016kjj}
H.~Afshar, D.~Grumiller, W.~Merbis, A.~Perez, D.~Tempo and R.~Troncoso,
  \emph{{Soft hairy horizons in three spacetime dimensions}},
  \href{https://doi.org/10.1103/PhysRevD.95.106005}{\emph{Phys. Rev.}
  {\bfseries D95} (2017) 106005},
  [\href{https://arxiv.org/abs/1611.09783}{{\ttfamily 1611.09783}}].

\bibitem{Afshar:2016uax}
H.~Afshar, D.~Grumiller and M.~M. Sheikh-Jabbari, \emph{{Near horizon soft hair
  as microstates of three dimensional black holes}},
  \href{https://doi.org/10.1103/PhysRevD.96.084032}{\emph{Phys. Rev.}
  {\bfseries D96} (2017) 084032},
  [\href{https://arxiv.org/abs/1607.00009}{{\ttfamily 1607.00009}}].

\bibitem{Afshar:2017okz}
H.~Afshar, D.~Grumiller, M.~M. Sheikh-Jabbari and H.~Yavartanoo, \emph{{Horizon
  fluff, semi-classical black hole microstates--- Log-corrections to BTZ
  entropy and black hole/particle correspondence}},
  \href{https://doi.org/10.1007/JHEP08(2017)087}{\emph{JHEP} {\bfseries 08}
  (2017) 087}, [\href{https://arxiv.org/abs/1705.06257}{{\ttfamily
  1705.06257}}].

\bibitem{Afshar:2018apx}
H.~Afshar, E.~Esmaeili and M.~M. Sheikh-Jabbari, \emph{{Asymptotic Symmetries
  in $p$-Form Theories}},
  \href{https://doi.org/10.1007/JHEP05(2018)042}{\emph{JHEP} {\bfseries 05}
  (2018) 042}, [\href{https://arxiv.org/abs/1801.07752}{{\ttfamily
  1801.07752}}].

\bibitem{Bunster:2010wv}
C.~Bunster and M.~Henneaux, \emph{{Can (Electric-Magnetic) Duality Be
  Gauged?}}, \href{https://doi.org/10.1103/PhysRevD.83.045031}{\emph{Phys.
  Rev.} {\bfseries D83} (2011) 045031},
  [\href{https://arxiv.org/abs/1011.5889}{{\ttfamily 1011.5889}}].

\end{thebibliography}\endgroup

\end{document}